\documentclass[twocolumn,preprintnumbers,amsmath,amssymb]{revtex4}
\usepackage{dcolumn}
\usepackage{bm}
\usepackage{graphicx}
\usepackage{amsmath}
\begin{document}

\title{Stability of spin dynamics for a spin-orbit coupled boson in a driven non-Hermitian double well}
\author{\ Zhida Luo$^{1}$, \ Yurui Yang$^{1}$, \ Jiaxi Cui$^{1}$, \ Wenjuan Li$^{2}$, \ Miaoqian Lu$^{1}$, \ Xinzhou Guan$^{1}$, \ Wenhua Hai$^{1}$, and Yunrong Luo$^{1}$\footnote{Corresponding author: lyr\underline{ }1982@hunnu.edu.cn}}
\affiliation{$^{1}$Key Laboratory of Low-dimensional Quantum Structures and Quantum Control of Ministry
of Education, Key Laboratory for Matter Microstructure and Function of Hunan Province, and Hunan Research Center of the Basic Discipline for Quantum Effects and Quantum Technologies, School of Physics and Electronics, Hunan Normal University, Changsha 410081, China\\
$^{2}$School of Physics and Electronic Information Engineering, Ningxia Normal University, Guyuan, Ningxia 756000, China}

\begin{abstract}

We study the stability of spin dynamics for a spin-orbit (SO) coupled boson held in a driven non-Hermitian double-well potential. Under high-frequency approximation, we analytically derive the Floquet states and complex Floquet quasienergies of the system and reveal a striking parity-dependent stability criterion: when the ratio of the Zeeman field strength to the driving frequency $\Omega/\omega$ is even, stable spin dynamics can be achieved for \emph{arbitrary} SO coupling strength. However, when $\Omega/\omega$ is odd, stability requires the SO coupling strength to be integer or half-integer values. Particularly, we find four types of stability boundary lines for non-zero bias field strength, in sharp contrast to the commonly observed stability regions. These results establish a tunable parity-governed mechanism for stabilizing spin dynamics in non-Hermitian cold atomic systems.

\end{abstract}

\maketitle

\section{Introduction}
In recent years, the study of stability in non-Hermitian physical systems has garnered significant attention, driven by the potential for novel phenomena and applications in quantum mechanics, optics, and condensed-matter physics\cite{moise, gana14, ashida, okuma, zhang}. Non-Hermitian systems, characterized by complex Hamiltonians, often exhibit unique behaviors such as exceptional points (EPs)\cite{du134, miri} and symmetry-breaking transitions\cite{chen134}, which can lead to the stabilization of such systems under certain conditions. For instance, periodic driving has been shown to stabilize non-Hermitian systems by rendering the eigenphases of the Floquet operator purely real, as demonstrated in the context of non-Hermitian Rabi models and superlattice potentials\cite{gong91}. This stabilization mechanism is attributed to the emergence of extended unitarity, where the system's dynamics remain coherent over multiple driving periods without exponential growth or decay. Recent advancements have also explored the role of parity-time ($\mathcal{PT}$) symmetry in stabilizing non-Hermitian systems\cite{luoxb94, luoxb95, zhou91}. $\mathcal{PT}$-symmetric systems can maintain real spectra under specific parameter regimes\cite{bender80, bender40}, and their stability has been investigated in various settings, including optical couplers\cite{luoxb94} and two-level systems\cite{luoxb95}. Moreover,the competition between gain and loss in non-Hermitian systems critically determines the stability of systems, such as the stability of quantum droplets in $\mathcal{PT}$-symmetric dual-core couplers\cite{zhou91} and the stable tunneling in an open double-well system\cite{xiao85}. These studies highlight the importance of balancing gain and loss to achieve stable dynamics.

Concurrently, the investigation of spin-orbit (SO) coupling in cold atom systems has been the focus of considerable research due to its potential for uncovering new quantum phenomena. The realization of SO coupling in ultracold atoms is typically achieved through the use of Raman lasers, which couple different hyperfine states of atoms, thereby inducing an effective SO coupling\cite{lin471}. This technique has enabled the study of various quantum effects, such as the spin-Hall effect\cite{kato306} and topological insulators\cite{bern314}, in a highly controllable manner.
Recent research has extended the study of SO coupling to non-Hermitian systems\cite{saka18, qin24, zhao108}, where gain and loss mechanisms are introduced to explore their impact on quantum dynamics. For instance, a control method was suggested for realizing full interband transitions in non-Hermitian SO coupled cold atomic systems by adding an extra non-Hermitian factor (atom loss) along with two-photon detuning, which allows flexible manipulation of quantum states and has potential uses in quantum simulations\cite{liu109}. The impact of gain and loss on the stability, phase transitions, and topological properties of SO coupled ultracold atoms in 2D optical lattices was also examined, revealing novel physical phenomena and offering theoretical and experimental support for the field\cite{xu65}. Additionally, research has shown that periodic driving can stabilize spin tunneling in a non-Hermitian SO coupled double-well system, countering the instability usually seen in non-Hermitian dynamics\cite{luo22}. Further investigation has looked into the spin Josephson effects in a non-Hermitian double well, showing that a net spin current can be sustained even with non-Hermitian elements\cite{tang55}. Exact analytical solutions have also been provided for a non-Hermitian double well with SO coupling under combined modulations, emphasizing the role of $\mathcal{PT}$ symmetry in the system's stability\cite{xie56}. However, in previous studies on non-Hermitian systems, parameters are usually chosen within a certain parameter region to ensure the dynamical stability of the system\cite{gong91, luo22, tang55, xie56}. Is it possible that for the system to be stable, the parameter values are not taken within a region but rather on a line?

In this paper, we theoretically investigate the stability of quantum spin dynamics for a SO coupled boson confined in a periodically driven non-Hermitian double well. In the high-frequency limit, we obtain analytical expressions for the Floquet states and the associated complex quasienergies and uncover an unexpected parity-dependent stability rule: when the ratio between the Zeeman field strength and the driving frequency $\Omega/\omega$ is even, suitable parameters can be found to stabilize the quantum spin dynamics of the system for arbitrary SO coupling strength. In contrast, when $\Omega/\omega$ is odd, the SO coupling strength can only take integer or half-integer values for the spin dynamics of the system to possibly be stable. For non-zero bias field strength, we find four distinct types of stability boundary lines, in contrast to the conventional region-based stability paradigm, thereby offering a new paradigm for stability conditions.

\section{Analytical solutions in the high-frequency approximation}

We consider a single SO coupled boson held in a driven non-Hermitian double well in which the dynamics is governed by a non-Hermitian Hamiltonian\cite{luo22, luoxb103, zou17}
\begin{eqnarray}\label{1}
\hat{H}(t)&=& -\nu (\hat{a}_{l}^{\dagger}e^{-i \pi \gamma \hat{\sigma}_{y}}\hat{a}_{r}+H.c.)+\frac{\Omega}{2}\sum_{j} (\hat{n}_{j \uparrow}-\hat{n}_{j\downarrow}) \nonumber\\ &+&\frac{\epsilon(t)}{2}\sum_{\sigma}(\hat{n}_{l \sigma}-\hat{n}_{r \sigma}).
\end{eqnarray}
Here, $\hat{a}_{j}^{\dag}=(\hat{a}_{j \uparrow}^{\dagger},\hat{a}_{j \downarrow}^{\dagger})$ and $\hat{a}_{j}=(\hat{a}_{j \uparrow },\hat{a}_{j \downarrow})^{T}$ ($T$ denotes the transpose) are matrices with elements representing the creation and annihilation operations of a pseudospin $\sigma=\uparrow, \downarrow$ boson in the $j$th $(j=l,r)$ well, respectively. $\hat{n}_{j \sigma}$ represents the number operator for pseudospin $\sigma$ in the $j$th well. $\nu$ denotes the tunneling rate between two wells without SO coupling, $\gamma$ is the SO coupling strength, $\hat{\sigma}_{y}$ is the {y}-component of the Pauli operator, $H.c.$ denotes the Hermitian conjugate of the preceding term. $\Omega$ is the effective Zeeman field strength. The form $\epsilon(t)=\epsilon\cos(\omega t)=(\epsilon_{1}+i \epsilon_{2})\cos(\omega t)$ has been introduced in prior studies\cite{gong91, zou17, li10}, where $\epsilon_{1}$ denotes the driving strength of bias field and $\epsilon_{2}$ denotes the gain-loss strength and $\omega$ is the driving frequency. In this work, we assume that $\epsilon_{1}\geq 0$ and $\epsilon_{2}> 0$.
Throughout this paper, $\hbar$ =1 and the parameters $\nu$, $\Omega$, $\epsilon_{1}$ , $\epsilon_{2}$ , $\omega$ are in units of the reference frequency $\omega_{0}=0.1 E_{R}$ with $E_R=k_{L}^2/2m=22.5$kHz being the single-photon recoil energy\cite{xue90}, and the time $t$ is normalized in units of $\omega_{0}^{-1}$\cite{luo93}. In experiment\cite{lin471}, the Zeeman field strength $\Omega$ is set as $-40 \omega_{0} \sim 40 \omega_{0}$, and the system parameters can be tuned over a wide range as follows\cite{xue90, luo93}: $\Omega, \epsilon_{1} \sim\omega \in [0, 100 \omega_{0}]$, and $\nu, \epsilon_{2} \sim\omega_{0}$.

Using the Fock basis $|0, \sigma\rangle$ $(|\sigma, 0\rangle)$ to represent the state where the spin $\sigma$ atom occupies the right (left) well and the left (right) well is empty, we can expand the quantum state of the SO coupled bosonic system as
\begin{eqnarray}\label{2}
|\Psi(t)\rangle &=& a_1(t)|0,\uparrow\rangle + a_2(t)|0,\downarrow\rangle + a_3(t)|\uparrow,0\rangle\nonumber\\ &+& a_4(t)|\downarrow,0\rangle,
\end{eqnarray}
where $a_{k}(t)$ ($k = 1,2,3,4$) represents the probability amplitude of the corresponding Fock state $|0,\sigma\rangle$ or $|\sigma,0\rangle$ (e.g, $a_{1}(t)$  represents the probability amplitude in Fock state $|0,\uparrow\rangle$), and the probability reads $P_{k}(t)=|a_{k}(t)|^{2}$.
Inserting equations (1) and (2) into the Schr\"{o}dinger equation $i\frac{\partial|\psi(t)\rangle}{\partial t}=\hat{H}(t)|\psi(t)\rangle$, we can obtain the coupled equations
\begin{eqnarray}\label{3}
i\dot{a}_1(t)&=&-\nu\cos(\pi\gamma)a_3(t)-\nu\sin(\pi\gamma)a_4(t)\nonumber\\&+&\frac{1}{2}(\Omega-\epsilon\cos(\omega t))a_1(t),\nonumber\\
i\dot{a}_2(t)&=&-\nu\cos(\pi\gamma)a_4(t)+\nu\sin(\pi\gamma)a_3(t)\nonumber\\&+&\frac{1}{2}(-\Omega-\epsilon\cos(\omega t))a_2(t),\nonumber\\
i\dot{a}_3(t)&=&-\nu\cos(\pi\gamma)a_1(t)+\nu\sin(\pi\gamma)a_2(t)\nonumber\\&+&\frac{1}{2}(\Omega+\epsilon\cos(\omega t))a_3(t),\nonumber\\
i\dot{a}_4(t)&=&-\nu\cos(\pi\gamma)a_2(t)-\nu\sin(\pi\gamma)a_1(t)\nonumber\\&+&\frac{1}{2}(-\Omega+\epsilon\cos(\omega t))a_4(t).
\end{eqnarray}

Due to the time-dependent coefficients, it is difficult to obtain the exact analytical solution of equation (3). Here, we try to find an approximate solution to equation (3) under high-frequency approximation where $\omega\gg\nu$\cite{luo22, luo110}. We introduce the slowly varying function $b_{k}(t)$ via the transformation
\begin{eqnarray}\label{4}
a_k(t)=b_k(t)X_{k},
\end{eqnarray}
where $k=1,2,3,4$, $X_{1,3}=e^{-i \int \frac{1}{2}[\Omega\mp\epsilon\cos(\omega t)]dt}$ and $X_{2,4}=e^{-i \int \frac{1}{2}[-\Omega\mp\epsilon\cos(\omega t)]dt}$. By employing the Fourier expansions $e^{\pm i \int \epsilon\cos(\omega t)dt}=\sum_{n=- \infty}^{\infty}{J}_{n}(\frac{\epsilon}{\omega})e^{\pm in\omega t}$ and $e^{\pm i \int [\epsilon\cos(\omega t)\pm\Omega]dt}=\sum_{n^{'}=- \infty}^{\infty}{J}_{n^{'}}(\frac{\epsilon}{\omega})e^{\pm i(n^{'}\pm\frac{\Omega}{\omega})\omega t}$, and neglecting the rapidly oscillating terms with $n\neq 0$ and $n^{'}\neq \mp\frac{\Omega}{\omega}$, we obtain a valid non-driven model
\begin{eqnarray}\label{5}
i\dot{b}_1(t)&= -J_{0}b_{3}(t)-J_{\frac{\Omega}{\omega}}b_{4}(t),\nonumber\\
i\dot{b}_2(t)&= -J_{0}b_{4}(t)+J_{-\frac{\Omega}{\omega}}b_{3}(t),\nonumber\\
i\dot{b}_3(t)&=-J_{0}b_{1}(t)+J_{-\frac{\Omega}{\omega}}b_{2}(t),\nonumber\\
i\dot{b}_4(t)&=-J_{0}b_{2}(t)-J_{\frac{\Omega}{\omega}}b_{1}(t).
\end{eqnarray}
Here the effective coupling constants are written as $J_{0}=\nu\cos(\pi\gamma) \mathcal{J}_{0}(\frac{\epsilon}{\omega})=\nu\cos(\pi\gamma) \mathcal{J}_{0}(\frac{\epsilon_{1}+i\epsilon_{2}}{\omega})$ and $J_{\pm\frac{\Omega}{\omega}}=\nu\sin(\pi\gamma) \mathcal{J}_{\pm\frac{\Omega}{\omega}}(\frac{\epsilon}{\omega})=\nu\sin(\pi\gamma) \mathcal{J}_{\pm\frac{\Omega}{\omega}}(\frac{\epsilon_{1}+i\epsilon_{2}}{\omega})$
with $\mathcal{J}_{n}(x)$ being the $n$-order ordinary Bessel function of $x$. When $\mathcal{J}_{0}(\frac{\epsilon}{\omega})$ and $\mathcal{J}_{\pm\frac{\Omega}{\omega}}(\frac{\epsilon}{\omega})$ are real values, if $\mid\mathcal{J}_{0}(\frac{\epsilon}{\omega})\mid\leq 1$ and $\mid\mathcal{J}_{\pm\frac{\Omega}{\omega}}(\frac{\epsilon}{\omega})\mid\leq 1$, the renormalized effective coupling is a general or conventional coupling in the high-frequency limit\cite{xue90, luo93}. Otherwise, if $\mid\mathcal{J}_{0}(\frac{\epsilon}{\omega})\mid> 1$ and/or $\mid\mathcal{J}_{\pm\frac{\Omega}{\omega}}(\frac{\epsilon}{\omega})\mid> 1$, the renormalized effective coupling is a special coupling, that is called strong coupling\cite{zou17}, which is hard to be implemented in a Hermitian system. However, it is very important in many applications\cite{longhi80, kfir123, fel124}.

\subsection{Floquet states and Floquet quasienergies}

Based on the Floquet theorem\cite{shirley138, sambe7}, it is known that for a time-periodic Hamiltonian model (1), the SO coupled ultracold atomic system exists Floquet solution and Floquet quasienergy. Therefore, the equation (2) which is the solution of the time-dependent Schr\"{o}dinger equation can be rewritten as the form $|\psi(t)\rangle=|\varphi(t)\rangle e^{-iE t}$, where $E$ is the Floquet quasienergy and
\begin{eqnarray}\label{6}
|\varphi(t)\rangle &=& A X_{1}|0,\uparrow\rangle+B X_{2}|0,\downarrow\rangle +C X_{3}|\uparrow,0\rangle\nonumber\\ &+& D X_{4}|\downarrow,0\rangle
\end{eqnarray}
is Floquet state which has the same period as the Hamiltonian (1). According to the relation between $a_k(t)$ and $b_k(t)$, the solutions of equation (5) can be constructed as $b_1(t)=A e^{-iEt}$, $b_2(t)=B e^{-iEt}$, $b_3(t)=C e^{-iEt}$, and $b_4(t)=D e^{-iEt}$, where $A$, $B$, $C$, and $D$ are constants and can be determined by the initial conditions. By substituting the stationary solutions into equation (5), the Floquet quasienergy $E_k$ and the corresponding constants $A_k, B_k, C_k, D_k$ for $k=1,2,3,4$ can be obtained as
\begin{eqnarray}\label{7}
B_{1,2}&=&-A_{1,2}\alpha_{\pm},C_{1,2}=\pm A_{1,2}\alpha_{\pm},D_{1,2}=\mp A_{1,2},\nonumber\\
B_{3,4}&=&-A_{3,4}\alpha_{\mp},C_{3,4}=\pm A_{3,4}\alpha_{\mp},D_{3,4}=\mp A_{3,4},\nonumber\\
E_{1,2}&=&\pm\frac{1}{2}\kappa_{\mp},E_{3,4}=\pm\frac{1}{2}\kappa_{\pm}.
\end{eqnarray}
Here the constants $\alpha_{\pm}=\frac{J_{\frac{\Omega}{\omega}}+J_{-\frac{\Omega}{\omega}}\pm\sqrt{4J_{0}^{2}+(J_{\frac{\Omega}{\omega}}+J_{-\frac{\Omega}{\omega}})^{2}}}{2J_{0}}$ and $\kappa_{\pm}=J_{\frac{\Omega}{\omega}}-J_{-\frac{\Omega}{\omega}}\pm\sqrt{4J_{0}^{2}+(J_{\frac{\Omega}{\omega}}+J_{-\frac{\Omega}{\omega}})^{2}}$. Such that the four Floquet states can be gotten as
\begin{eqnarray}\label{8}
|\varphi_{k}(t)\rangle&=A_{k}X_{1}|0,\uparrow\rangle+B_{k}X_{2}|0,\downarrow\rangle\nonumber\\
&+C_{k}X_{3}|\uparrow,0\rangle+D_{k}X_{4}|\downarrow,0\rangle.
\end{eqnarray}

\subsection{General coherent non-Floquet state}

In order to study the stability of  the system's quantum spin dynamics, we have to consider the linear superposition of Floquet states\cite{jinas322}, and the linear superposition state is called the non-Floquet state\cite{lu83}. It can be constructed by the form
\begin{eqnarray}\label{9}
|\psi(t)\rangle&=&\sum^{4}_{k=1}f_k|\varphi_k (t)\rangle e^{-iE_k t}\nonumber\\
&=&X_{1}d_{1}(t)|0,\uparrow\rangle +X_{2}d_{2}(t)|0,\downarrow\rangle+X_{3}d_{3}(t)|\uparrow,0\rangle\nonumber \\ &+&X_{4}d_{4}(t)|\downarrow,0\rangle,
\end{eqnarray}
where $f_k$ is the superposition coefficient determined by the initial conditions, and the probability amplitudes are renormalized as
$d_1(t)=\sum^{4}_{k=1}f_k A_k e^{-iE_kt}$, $d_2(t)=\sum^{4}_{k=1}f_k B_k e^{-iE_kt}$, $d_3(t)=\sum^{4}_{k=1}f_k C_k e^{-iE_kt}$, and $d_4(t)=\sum^{4}_{k=1}f_k D_k e^{-iE_kt}$. The general non-Floquet state (9) implies quantum interference between four Floquet states with different quasienergies, which will lead to the enhancement or suppression of spin quantum tunneling of this system.

\section{Stability analysis of spin dynamics for even and odd $\Omega$/$\omega$}

It is well known that for a non-Hermitian system the stability of this system strongly depends on the imaginary part of the complex quasienergy. From equation (7), we can see the Floquet quasienergy $E_k$ is determined by the renormalized effective coupling constants $J_{0}$ and $J_{\pm\frac{\Omega}{\omega}}$. For the convenience of discussion, the complex Floquet quasienergy $E_k$ is written in this form $E_{k}$$=$Re($E_{k}$)+i Im($E_{k}$) ($k=1,2,3,4$, Re and Im represent the real part and imaginary part of a complex number respectively). Based on the stability criterion presented in our previous work, see Ref.\cite{luo22}, we know that a non-Hermitian system is stable in the following two cases.

\emph{Case 1}. When all of Im($E_{k}$) are equal to zero, in other words, the Floquet quasienergies are all real, the system is stable and the time-evolutions of all probabilities are periodic. This usually occurs when the gain-loss coefficients is balanced.

\emph{Case 2}. When some of Im($E_{k}$) are equal to zero and the others of Im($E_{k}$) are less than zero, the system is also stable and the time-evolutions of all probabilities tend to constants at $t\rightarrow\infty$. This generally happens under unbalanced gain and loss.

In this paper, due to the fact that the gain-loss coefficients is balanced, thus, we will present the system stability analysis by employing the stability criterion of case 1.

However, we also notice that the Floquet quasienergy $E_{k}$ is related to the Bessel functions $\mathcal{J}_{0}(\frac{\epsilon_{1}+i\epsilon_{2}}{\omega})$ and $\mathcal{J}_{\pm\frac{\Omega}{\omega}}(\frac{\epsilon_{1}+i\epsilon_{2}}{\omega})$ with complex variables, which is distinguished as the following two scenarios.

Scenario (i). When $\epsilon_{1}= 0$ and $\epsilon_{2}> 0$, this system is $\mathcal{PT}$-symmetric\cite{xie56, guo103}. At this moment, $\mathcal{J}_{0}(\frac{\epsilon_{1}+i\epsilon_{2}}{\omega})=\mathcal{J}_{0}(\frac{i\epsilon_{2}}{\omega})> 1$, and $\mathcal{J}_{\pm\frac{\Omega}{\omega}}(\frac{\epsilon_{1}+i\epsilon_{2}}{\omega})=\mathcal{J}_{\pm\frac{\Omega}{\omega}}(\frac{i\epsilon_{2}}{\omega})$ is real for even $\Omega$/$\omega$. In this case, all the Floquet quasienergies $E_{k}$ are real, and the system is stable. While for odd $\Omega$/$\omega$, $\mathcal{J}_{\pm\frac{\Omega}{\omega}}(\frac{i\epsilon_{2}}{\omega})$ is a pure imaginary number and
the stability of the system depends on the SO coupling strength.

Scenario (ii). When $\epsilon_{1}> 0$ and $\epsilon_{2}> 0$, this system is not $\mathcal{PT}$-symmetric. $\mathcal{J}_{0}(\frac{\epsilon_{1}+i\epsilon_{2}}{\omega})$ and $\mathcal{J}_{\pm\frac{\Omega}{\omega}}(\frac{\epsilon_{1}+i\epsilon_{2}}{\omega})$ are generally complex numbers. But, when $\epsilon_{1}/\omega$ and $\epsilon_{2}/\omega$ take appropriate values, $\mathcal{J}_{0}(\frac{\epsilon_{1}+i\epsilon_{2}}{\omega})$ and $\mathcal{J}_{\pm\frac{\Omega}{\omega}}(\frac{\epsilon_{1}+i\epsilon_{2}}{\omega})$ may be real, that leads to the Floquet quasienergies $E_{k}$ being real values. Such that the system can still maintain stability.

Next, we will discuss the stability of the system separately for $\Omega$/$\omega$ being odd and even, as the quasienergy $E_{k}$ depends on the parity of $\Omega$/$\omega$.

\subsection{Even $\Omega$/$\omega$}

When $\Omega$/$\omega$ takes even values, from equation (7), we can obtain $E_{1}=E_{2}=-E_{3}=-E_{4}=-\rho$ with $\rho=\sqrt{J_{0}^{2}+J_{\frac{\Omega}{\omega}}^{2}}$. It is evident that the stability of the system depends on Im($\rho$). Below, we will discuss the impact of different parameter values on the dynamical stability of the system separately in scenario (i) and scenario (ii).

1. Scenario (i)

For even $\Omega$/$\omega$ and $\epsilon_{1}= 0$ and $\epsilon_{2}> 0$, $\mathcal{J}_{0}(\frac{i\epsilon_{2}}{\omega})$ and $\mathcal{J}_{\pm\frac{\Omega}{\omega}}(\frac{i\epsilon_{2}}{\omega})$ are real values, and $\mathcal{J}_{0}(\frac{i\epsilon_{2}}{\omega})> 1$ and $\mid\mathcal{J}_{\pm\frac{\Omega}{\omega}}(\frac{i\epsilon_{2}}{\omega})\mid> 0$. Therefore, from equation (5), it can be seen that the coupling between $b_1$ ($b_2$) and $b_3$ ($b_4$) is the strong coupling, which means that the coupling between state $|0,\sigma\rangle$ and state $|\sigma,0\rangle$ is the strong coupling, namely, the quantum tunneling without spin-flipping is enhanced, which helps to suppress the decoherence effect of non-spin-flipping tunneling\cite{uch68}. However, whether the coupling between $b_1$ ($b_2$) and $b_4$ ($b_3$) or between state $|0,\sigma\rangle$ and state $|\sigma',0\rangle$ ($\sigma$ and $\sigma'$ denote different spin directions) is strong depends on whether $\mid\mathcal{J}_{\pm\frac{\Omega}{\omega}}(\frac{i\epsilon_{2}}{\omega})\mid> 1$ or $\mid\mathcal{J}_{\pm\frac{\Omega}{\omega}}(\frac{i\epsilon_{2}}{\omega})\mid\leq 1$. If $\mid\mathcal{J}_{\pm\frac{\Omega}{\omega}}(\frac{i\epsilon_{2}}{\omega})\mid> 1$, the coupling between state $|0,\sigma\rangle$ and state $|\sigma',0\rangle$ is the strong coupling. Otherwise, the coupling between them is the conventional coupling.

To further understand the impact of the gain-loss factor $\epsilon_{2}$ on the coupling (or quantum tunneling rate), we introduce the fidelity $F(t)$ that is the square overlap of the initial state $|\psi(t=0)\rangle$ and the evolving state $|\psi(t)\rangle$, namely, $F(t)=|\langle\psi(t=0)|\psi(t)\rangle|^{2}$. Here, we set the initial state $|0,\uparrow\rangle$ to plot the time evolutions of quantum fidelity $F(t)$ for different gain-loss coefficients $\epsilon_{2}$ in figure 1. In figure 1 (a), the SO coupling strength $\gamma=0.5$ means the initial state $|0,\uparrow\rangle$ is only coupled with the state $|\downarrow,0\rangle$ and the effective coupling factor is $J_{\pm\frac{\Omega}{\omega}}$. When $\epsilon_{2}=100$ and $\epsilon_{2}=114$, $\mid\mathcal{J}_{\pm\frac{\Omega}{\omega}}(\frac{i\epsilon_{2}}{\omega})\mid=\mid\mathcal{J}_{\pm2}(2i)\mid=0.6889<1$ and $\mid\mathcal{J}_{\pm\frac{\Omega}{\omega}}(\frac{i\epsilon_{2}}{\omega})\mid=\mid\mathcal{J}_{\pm2}(\frac{114i}{50})\mid=0.9812\approx 1$, respectively, which mean the coupling between states $|0,\uparrow\rangle$ and $|\downarrow,0\rangle$ is the conventional coupling. When $\epsilon_{2}=150$, $\mid\mathcal{J}_{\pm\frac{\Omega}{\omega}}(\frac{i\epsilon_{2}}{\omega})\mid=\mid\mathcal{J}_{\pm2}(3i)\mid=2.2452>1$ means the strong coupling between states $|0,\uparrow\rangle$ and $|\downarrow,0\rangle$ occurs. It is evident from Figure 1 (a) that the fidelity decays more rapidly in the strong coupling case compared to the conventional coupling case. In figure 1 (b), the SO coupling strength $\gamma=1$ means the initial state $|0,\uparrow\rangle$ is only coupled with the state $|\uparrow,0\rangle$ and the effective coupling factor is $J_{0}$. Since $\mathcal{J}_{0}(\frac{i\epsilon_{2}}{\omega})$ is always greater than 1 for $\epsilon_{2}>0$, thus, the quantum tunneling without spin-flipping between state $|0,\uparrow\rangle$ and state $|\uparrow,0\rangle$ is always enhanced. By comparing figure 1 (a) and figure 1 (b), it is not difficult to see that for the same gain-loss factor $\epsilon_{2}$ , the quantum tunneling without spin-flipping is faster than that with spin-flipping.

\begin{figure}[htp]\center
\includegraphics[height=1.3in,width=1.6in]{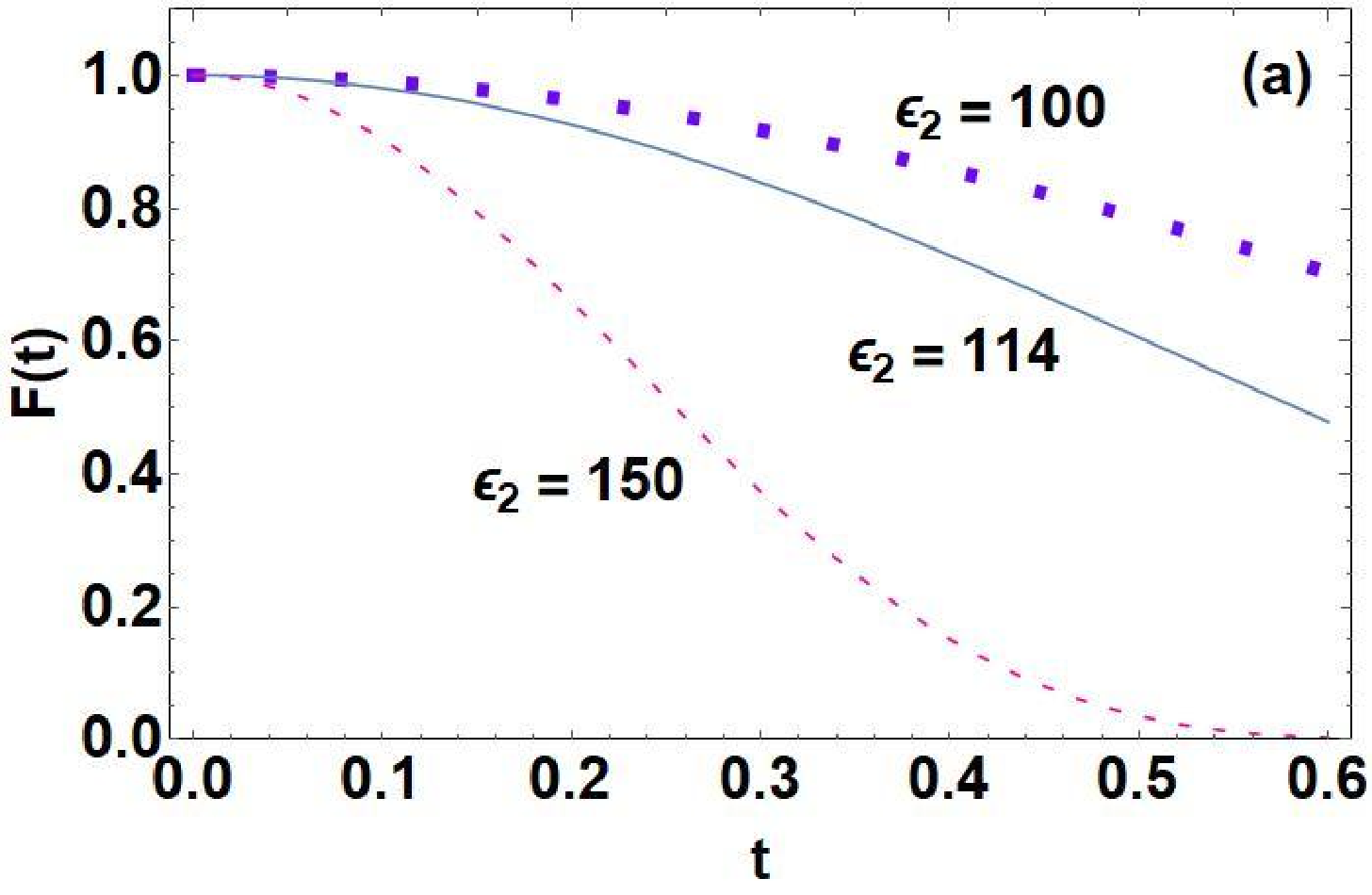}
\includegraphics[height=1.3in,width=1.6in]{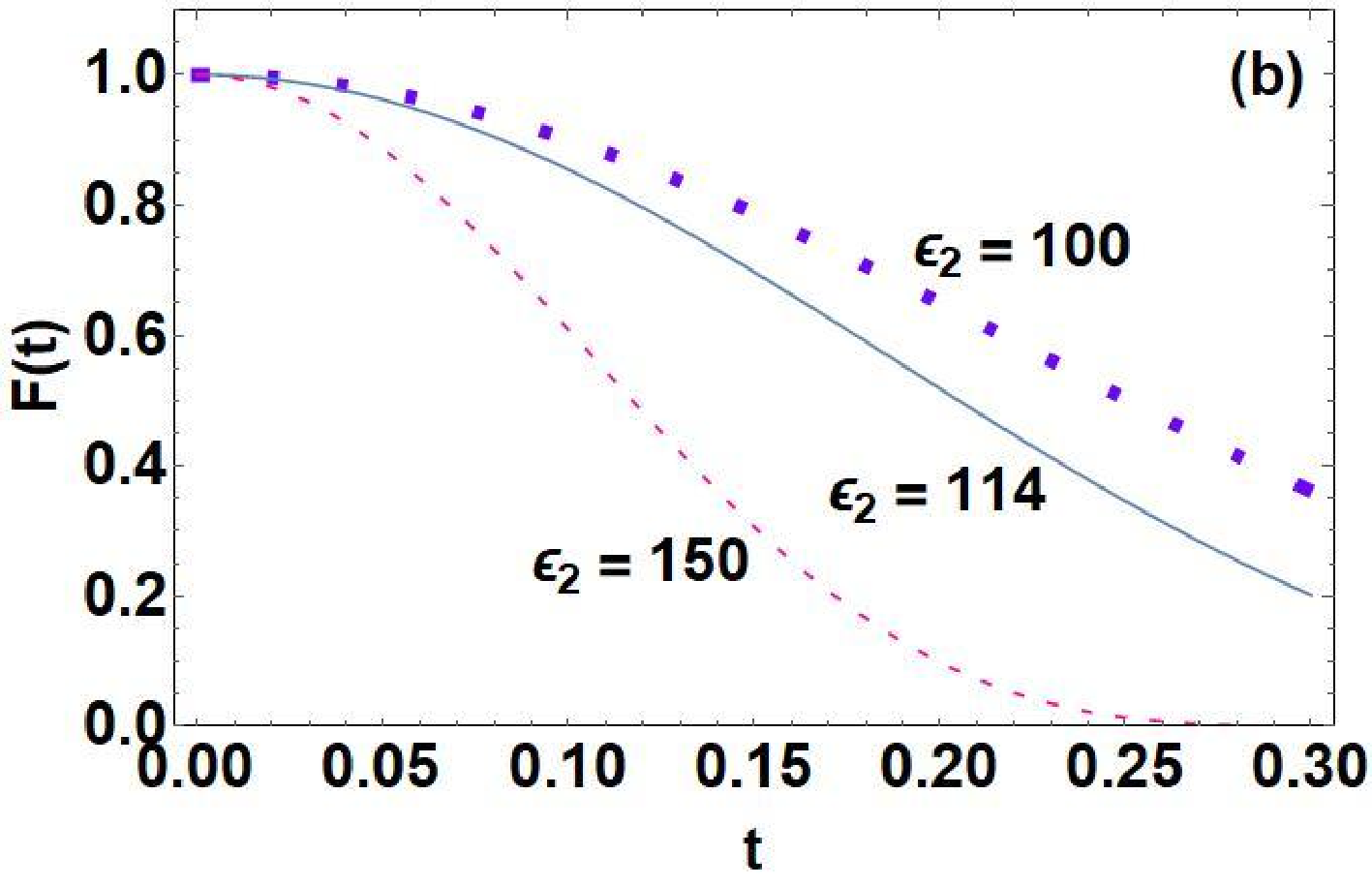}
\caption{\scriptsize{The time evolutions of the quantum fidelity $F(t)$ between the initial state $|\psi(t=0)\rangle=|0,\uparrow\rangle$ and the evolving state $|\psi(t)\rangle$ for $\epsilon_{2}=100$ (dotted line), $\epsilon_{2}=114$ (solid line), $\epsilon_{2}=150$ (dashed line), respectively. The other parameters are chosen as $\epsilon_{1}=0$, $\omega=50$, $\Omega=100$, $\nu=1$ for (a)$\gamma=0.5$ and (b)$\gamma=1$. All parameters adopted in these figures are dimensionless.}}.
\end{figure}

2. Scenario (ii)

For even $\Omega$/$\omega$ and $\epsilon_{1}> 0$ and $\epsilon_{2}> 0$, only when $\epsilon_{1}$ and $\epsilon_{2}$ take appropriate values, $\rho$ is real, namely, Im($\rho$) is equal to zero, that leads to the system being stable. Thus, in order to comprehensively investigate the impact of system parameters on system stability, we have plotted the functional relationships between Im($\rho$) and $\epsilon_1$/$\omega$, $\epsilon_2$/$\omega$, and $\gamma$ as shown in figures 2-4, where the red lines are boundary lines between Im($\rho$)$>0$ and Im($\rho$)$<0$. Based on boundary line characteristics, we classify them into four types:

Type I. Discontinuous boundary line. When the parameters are taken on this boundary line, Im($\rho$) is not equal to zero and the system is unstable.

Type II. Continuous boundary line. When the parameters are taken on this boundary line, Im($\rho$) is equal to zero and the system is stable.

Type III. Continuous boundary line with gaps. When the parameters are taken on the continuous part of this type of boundary line, Im($\rho$) is equal to zero and the system is stable. When the parameters are taken at the gap, Im($\rho$) is not equal to zero and the system is unstable.

Type IV. Composite boundary line, which consists of continuous boundary line and discontinuous boundary line. When the parameters are taken on the continuous part of this type of boundary line, Im($\rho$) is equal to zero and the system is stable. When the parameters are taken on the discontinuous part of this type of boundary line, Im($\rho$) is not equal to zero and the system is unstable.

\begin{figure}[htbp]\centering
\includegraphics[height=1.8in,width=1.8in]{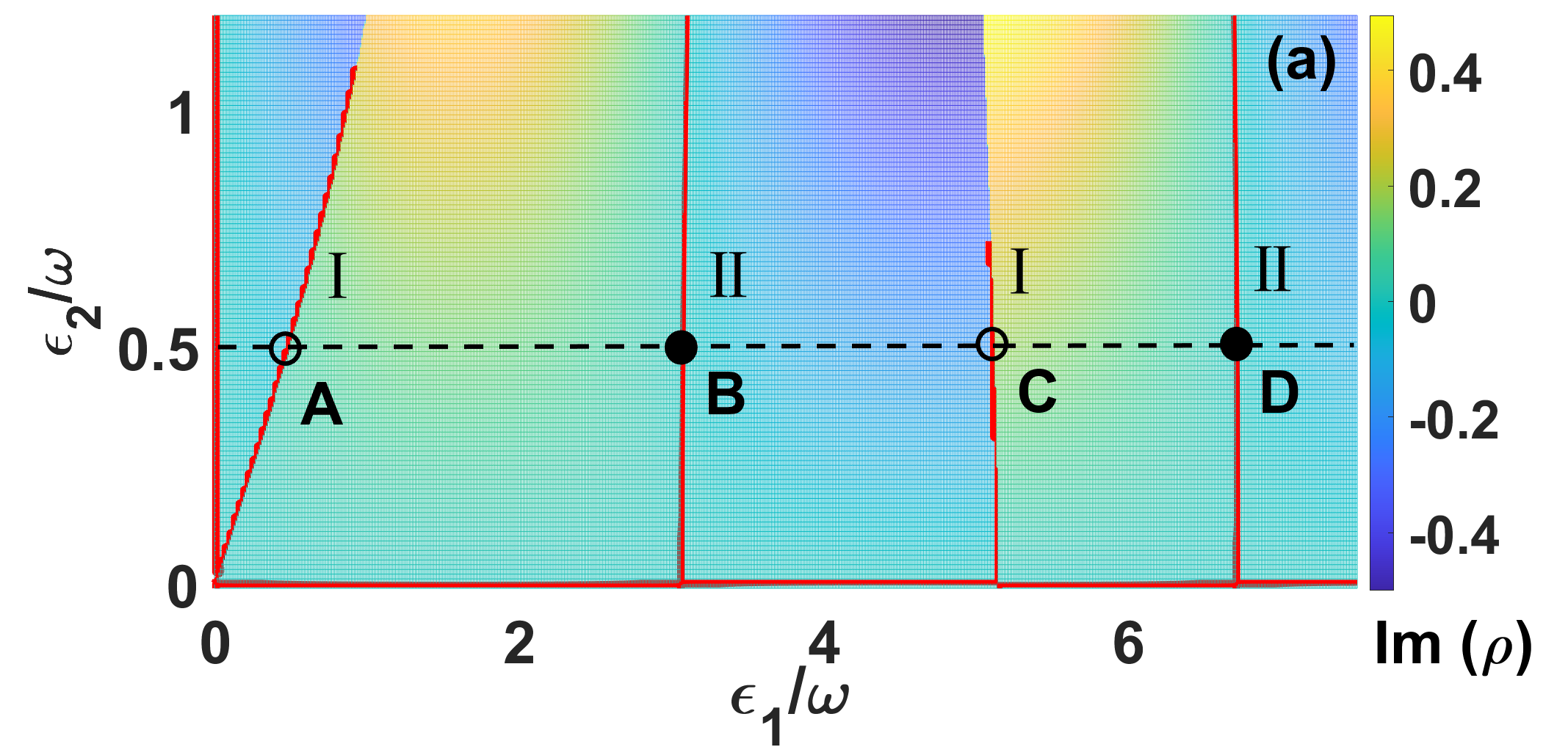}
\includegraphics[height=1.7in,width=1.5in]{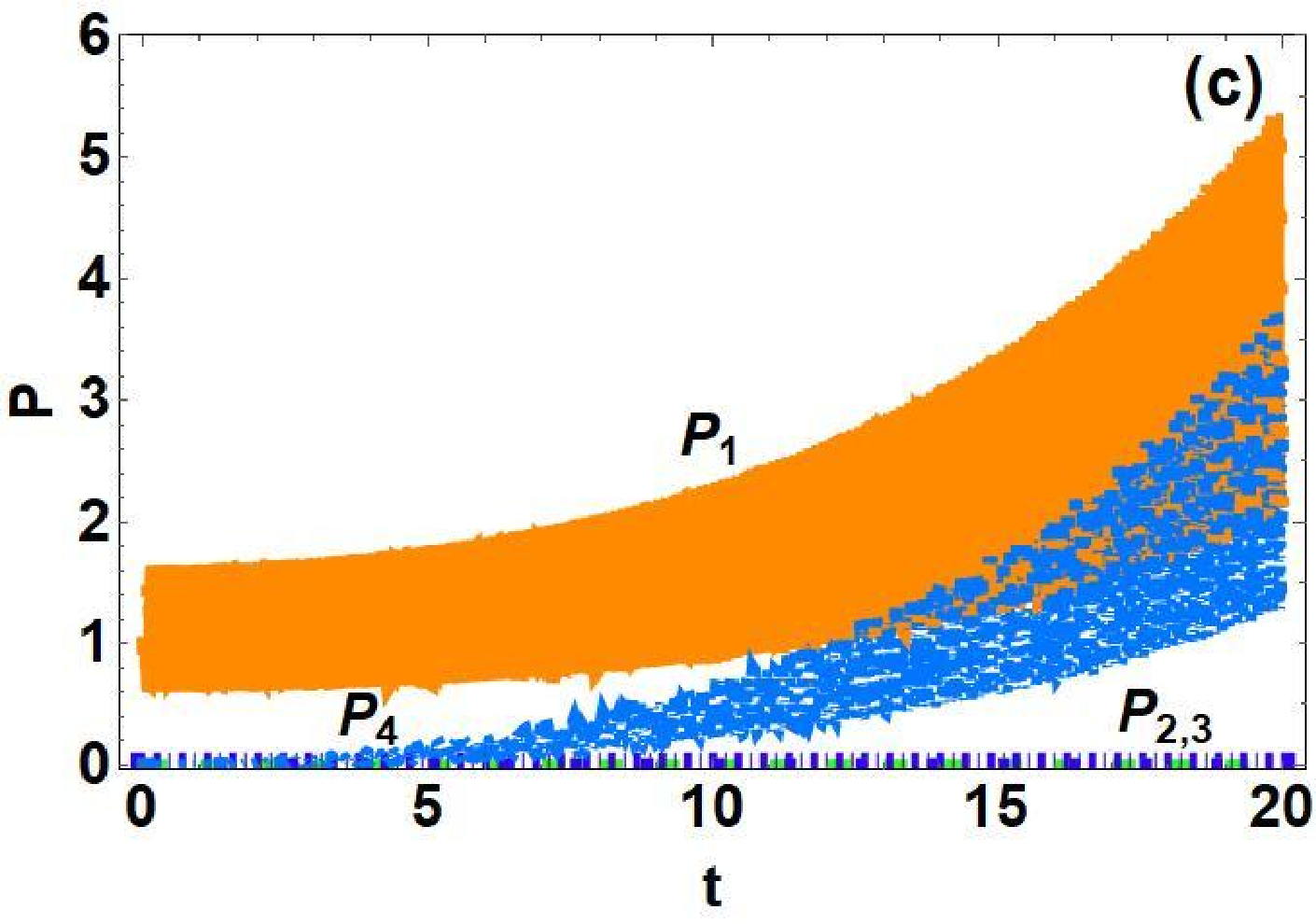}
\includegraphics[height=1.8in,width=1.8in]{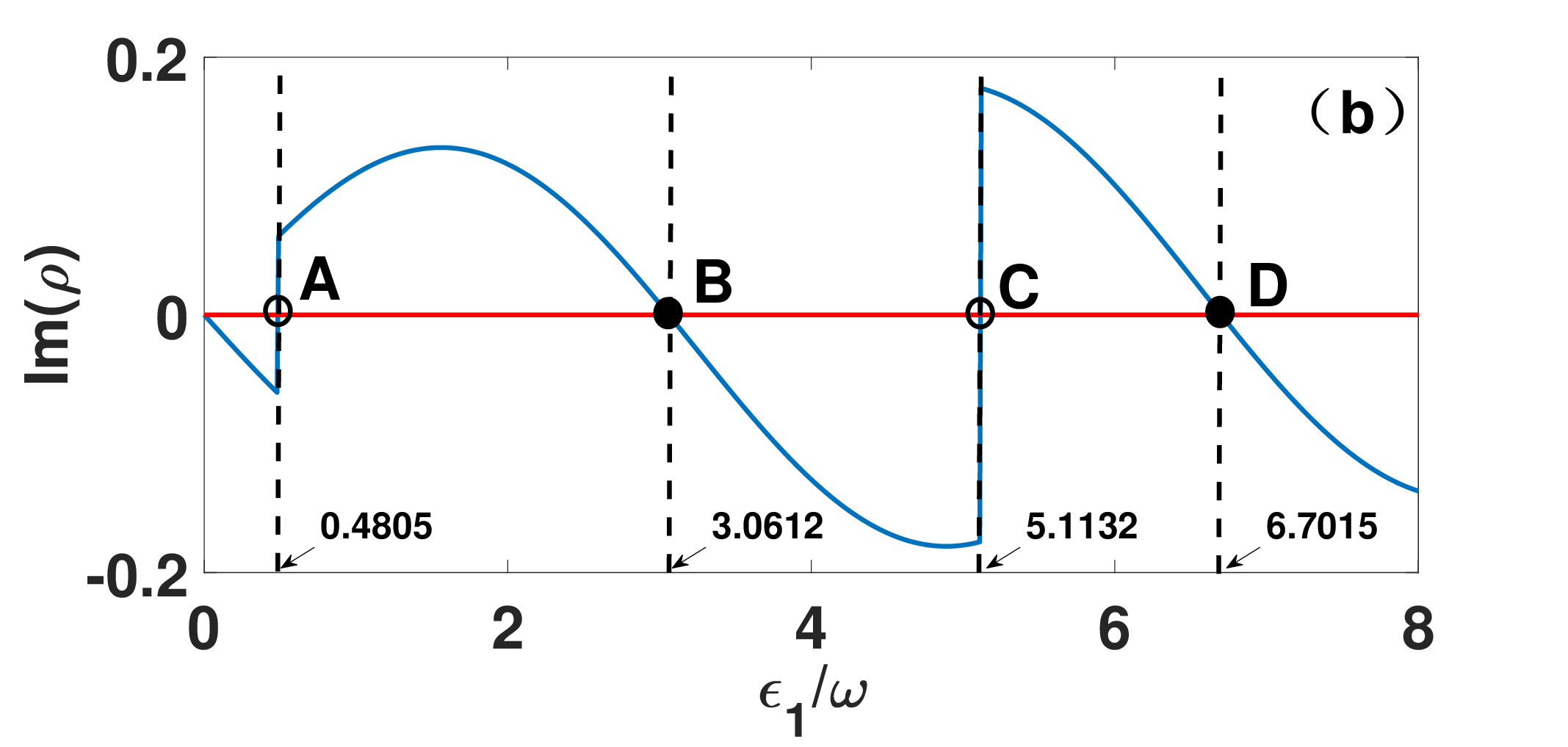}
\includegraphics[height=1.7in,width=1.5in]{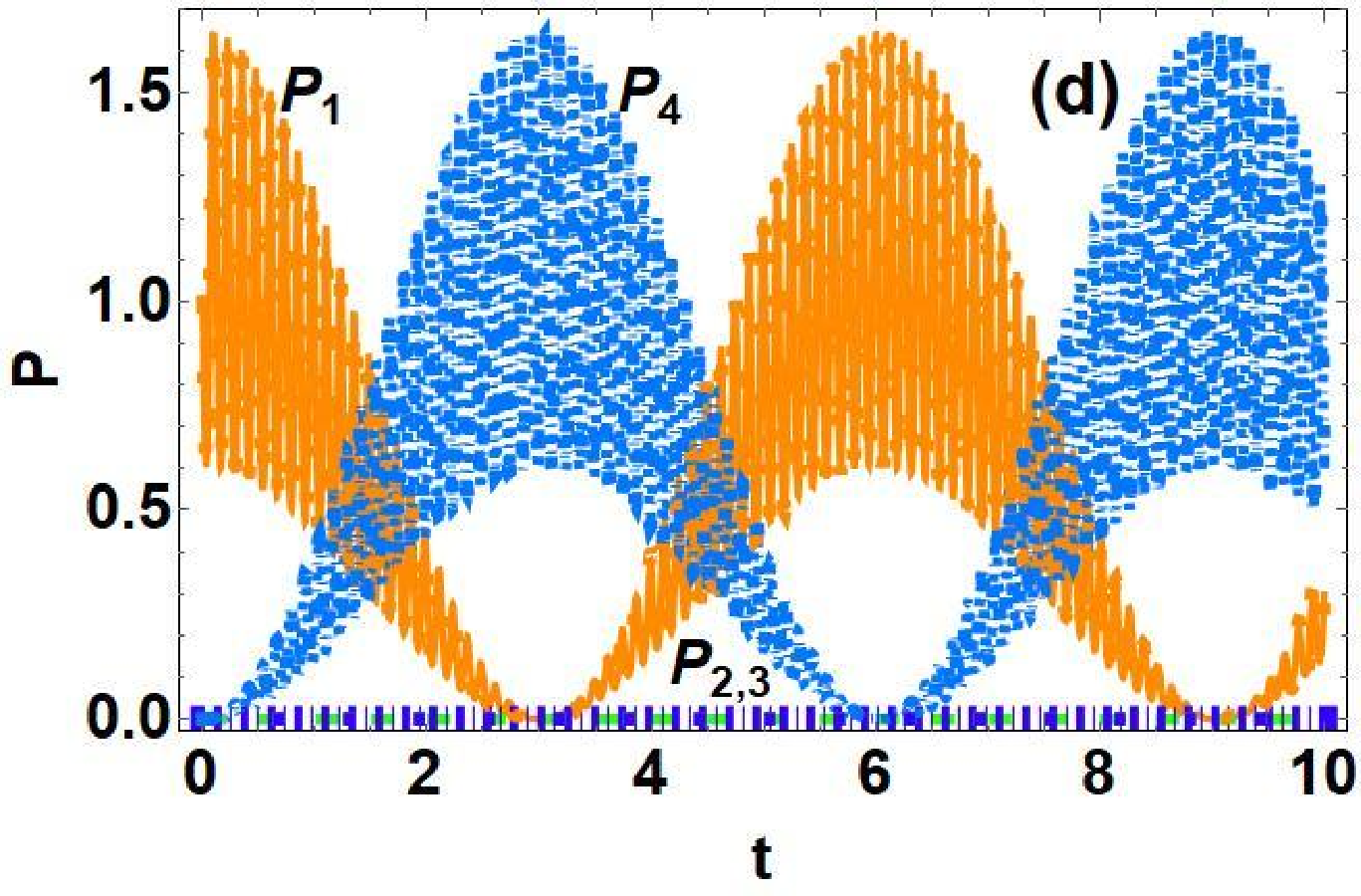}
\caption{\scriptsize{(a) Im($\rho$) as a function of $\frac{\epsilon_{1}}{\omega}$ and $\frac{\epsilon_{2}}{\omega}$. (b)Im($\rho$) as a function of $\frac{\epsilon_{1}}{\omega}$ for $\frac{\epsilon_{2}}{\omega}=0.5$. (c) and (d) show the time-evolution curves of probabilities $P_{k}=|a_{k}|^{2}$ for (c)$\frac{\epsilon_1}{\omega}=0.4805$ and $\frac{\epsilon_{2}}{\omega}=0.5$, and (d)$\frac{\epsilon_1}{\omega}=3.0612$ and $\frac{\epsilon_{2}}{\omega}=0.5$,  starting the system with a spin-up particle in the right well. The other parameters are chosen as $\gamma=0.5$, $\nu=1$, $\Omega=100$, and $\omega=50$. Shown in figures 2(c)and 2(d) are the numerical results from the origin model (1), and the dots denote the analytical correspondences obtained from the effective model (5). }}
\end{figure}

In order to verify the characteristics of the distinct types of boundary lines, we perform the following detailed analysis.

In figure 2(a), we set the parameters $\Omega=100$, $\omega=50$, $\nu=1$, $\gamma=0.5$, and plot Im($\rho$) as a function of $\frac{\epsilon_{1}}{\omega}$ and $\frac{\epsilon_{2}}{\omega}$. It can be seen that there are two types of boundary lines in figure 2(a), namely, type I and type II. What are the differences between them? When the parameters are taken on these two types of boundary lines,
is Im($\rho$) equal to 0? To answer this question, we arbitrarily select a value of $\frac{\epsilon_2}{\omega}$ (e.g., $\frac{\epsilon_2}{\omega}=0.5$) and draw a horizontal dashed line, which intersects the type I and type II boundary lines at four points, A, B, C, and D, as shown in figure 2(a). Then, we set $\frac{\epsilon_2}{\omega}=0.5$ and plot Im($\rho$) as a function of
$\frac{\epsilon_{1}}{\omega}$, see figure 2(b), the other parameters are the same as in figure 2(a). It can be easily seen from figure 2(b) that points A and C on the discontinuous boundary line of type I are jump points, with the corresponding Im($\rho$) not equal to 0, and the system is unstable. In contrast, points B and D on the continuous boundary line of type II are continuous points, with the corresponding Im($\rho$) equal to 0, and the system is stable. To verify our analysis, as an example, we select the initial state of the system to be state $\mid 0,\uparrow\rangle$, and take
$\frac{\epsilon_1}{\omega}=0.4805$ corresponding to the jump point A and $\frac{\epsilon_1}{\omega}=3.0612$ corresponding to the continuous point B in figure 2(b) to plot the evolution curves of probability over time, as shown in figures 2(c) and 2(d). As can be seen from figures 2(c) and 2(d), when the parameters are taken on the discontinuous boundary line of type I, the probabilities $P_1$ and $P_4$ increase exponentially, indicating that the system is unstable; when the parameters are taken on the continuous boundary line of type II, the probabilities $P_1$ and $P_4$ exhibit periodic oscillations, namely, the generalized Rabi oscillation between state $\mid 0,\uparrow\rangle$ and state $\mid \downarrow,0\rangle$ occurs\cite{gong91,luo22}, indicating that the system is stable. This confirms that our analysis is correct.

\begin{figure}[htp]\center
\includegraphics[height=1.8in,width=1.8in]{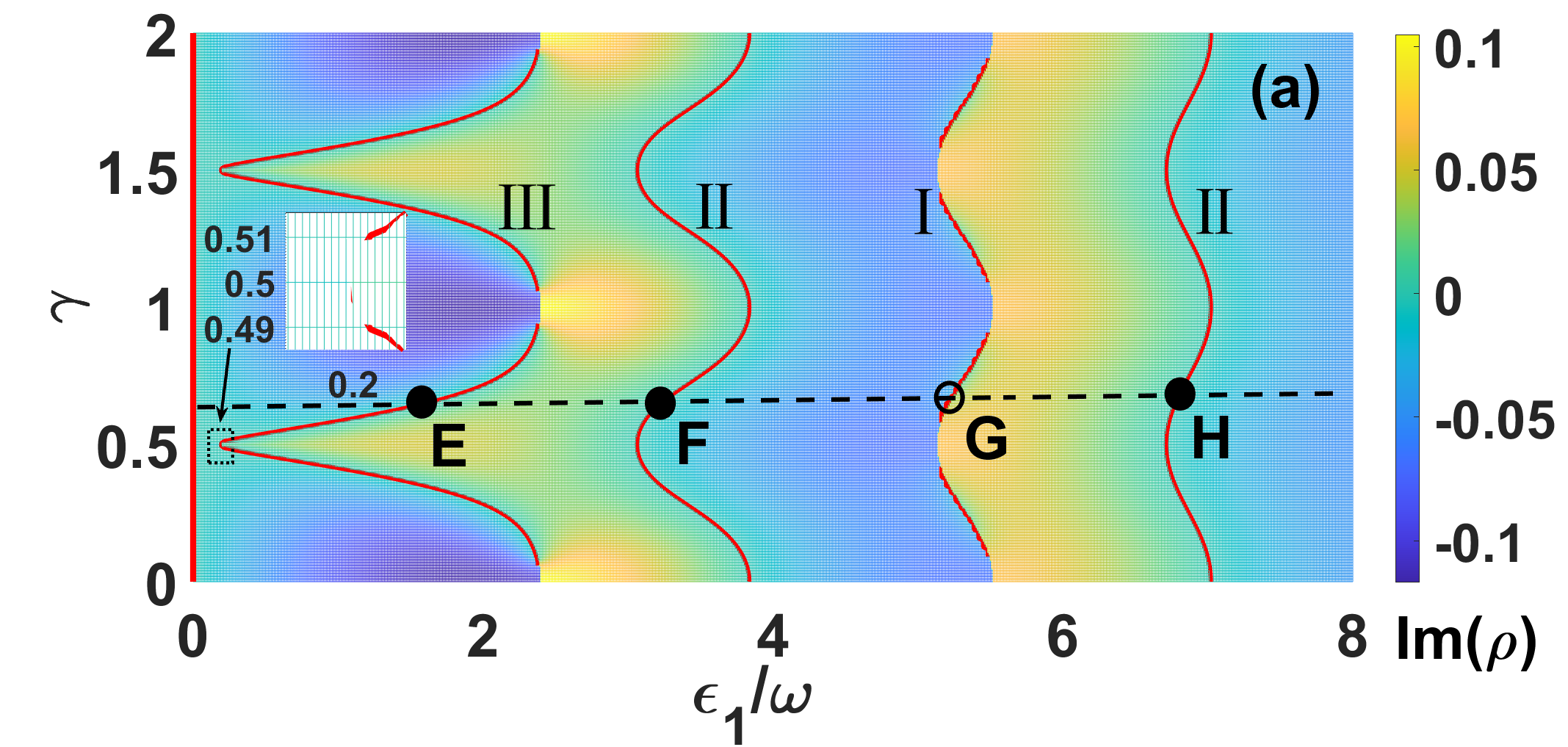}
\includegraphics[height=1.7in,width=1.5in]{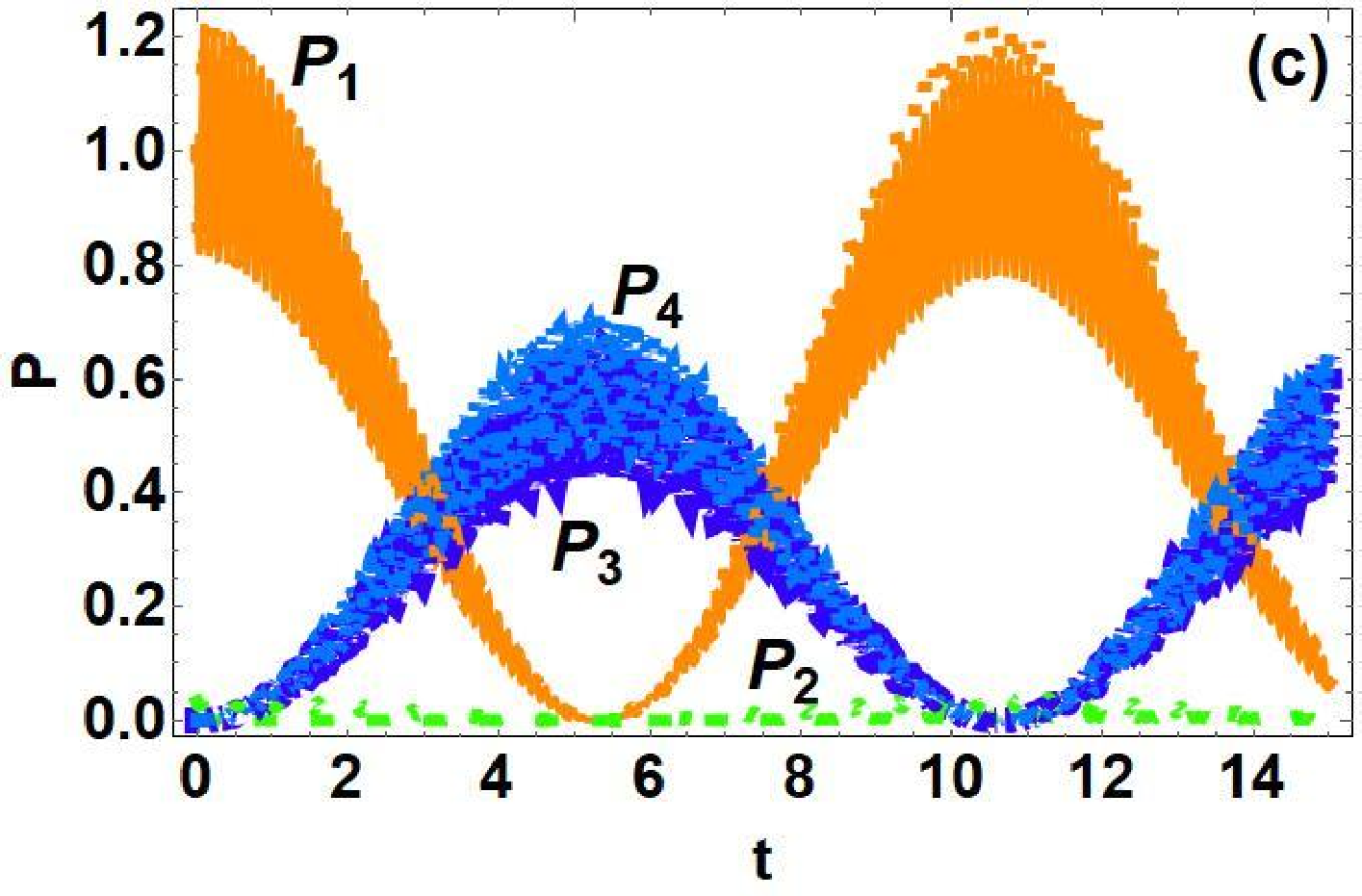}
\includegraphics[height=1.8in,width=1.8in]{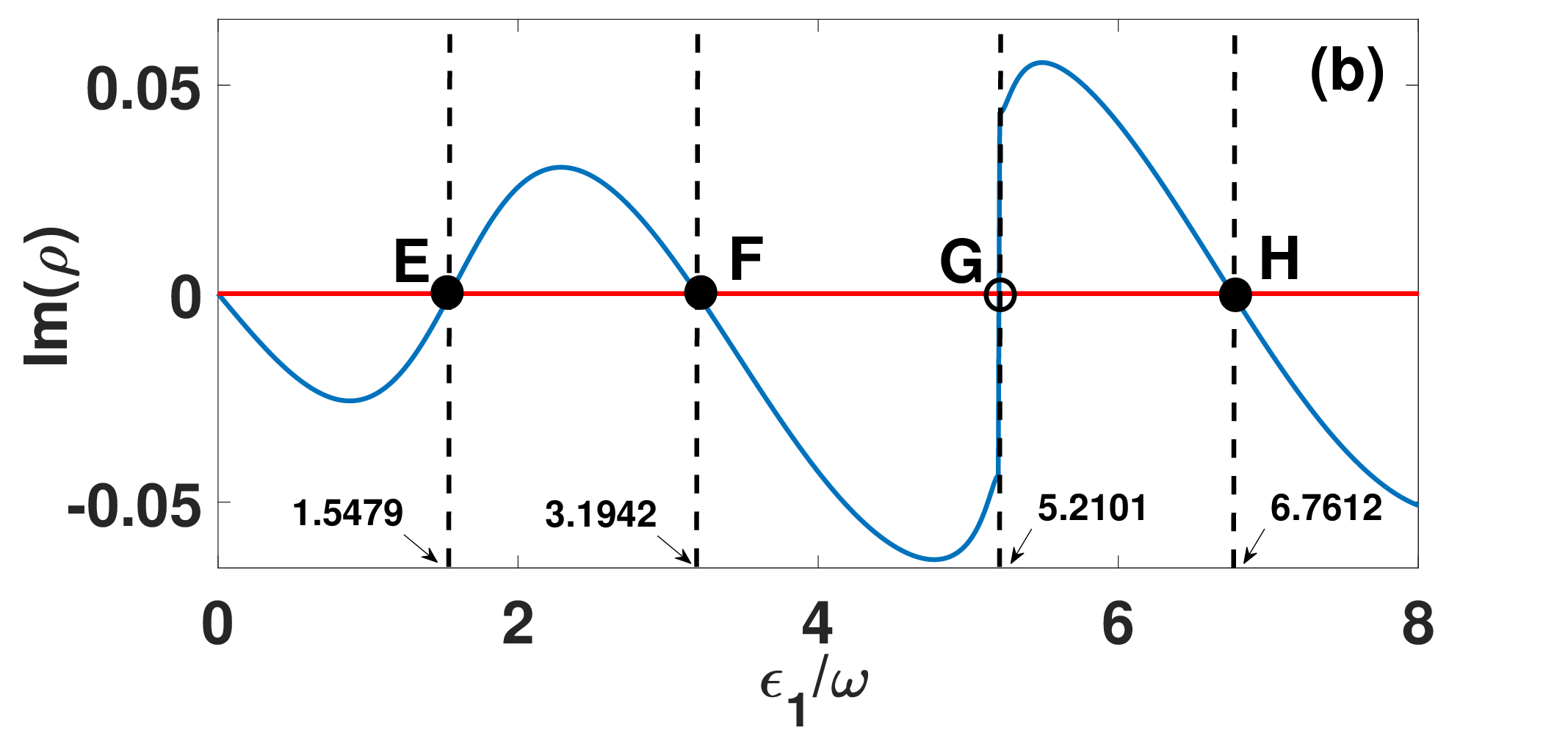}
\includegraphics[height=1.7in,width=1.5in]{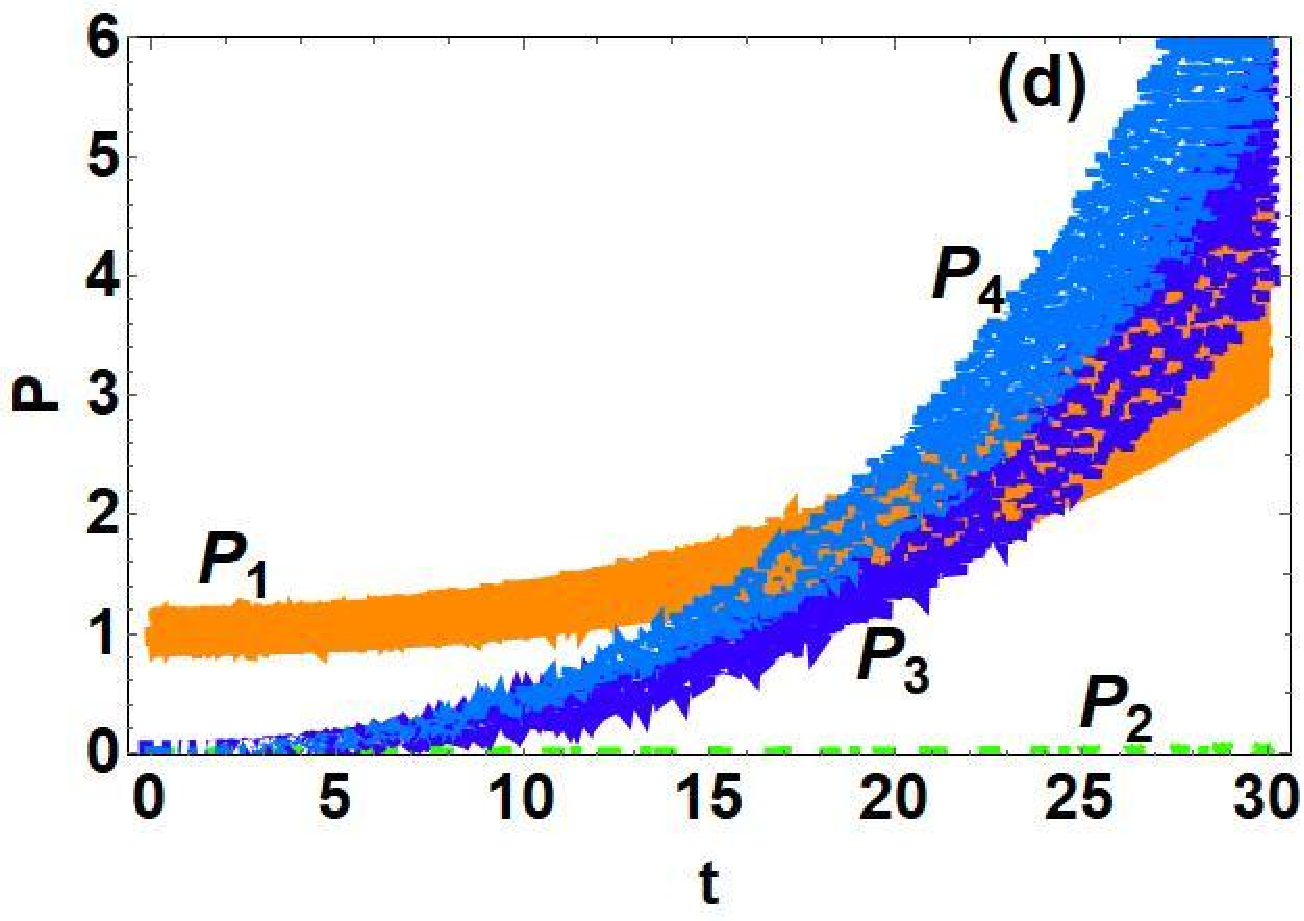}
\caption{\scriptsize{(a) Im($\rho$) as a function of $\frac{\epsilon_{1}}{\omega}$ and $\gamma$. (b) Im($\rho$) as a function of $\frac{\epsilon_{1}}{\omega}$ for $\gamma=0.64$. (c) and (d) show the time-evolution curves of probabilities $P_{k}=|a_{k}|^{2}$ for (c)$\frac{\epsilon_{1}}{\omega}=1.5479$ and $\gamma=0.64$, and (d)$\frac{\epsilon_{1}}{\omega}=5.2101$ and $\gamma=0.64$, starting the system with a spin-up particle in the right well. The other parameters are chosen as $\frac{\epsilon_{2}}{\omega}=0.2$, $\nu=1$, $\Omega=100$, and $\omega=50$.}}
\end{figure}

In figure 3(a), we set the parameters $\Omega=100$, $\omega=50$, $\nu=1$, $\frac{\epsilon_2}{\omega}=0.2$, and plot Im($\rho$) as a function of $\frac{\epsilon_{1}}{\omega}$ and $\gamma$. We find that there are three types of boundary lines in figure 3(a), namely, type I, type II and type III. Then, what is the difference between the type III boundary line and the type I and type II boundary lines? Similar to figure 2, we arbitrarily select a value of $\gamma$, such as $\gamma=0.64$ as an example to draw a horizontal dashed line, which intersects the three types of boundary lines at four points E, F, G, and H, respectively, as shown in figure 3(a). Then, we take $\gamma=0.64$ and plot Im($\rho$) as a function of $\frac{\epsilon_{1}}{\omega}$, see figure 3(b).
It can be seen from figure 3(b) that point E on the type III boundary line is similar to points F and H on the type II boundary line in that they are all continuous points with Im($\rho$)$=$0, and the system is stable. The point G on the type I discontinuous boundary line is a jump point, with the corresponding Im($\rho$)$\neq$0, and the system is unstable. As an example,
in figures 3(c) and 3(d), we take $\frac{\epsilon_1}{\omega}=1.5479$ corresponding to the continuous point E and $\frac{\epsilon_1}{\omega}=5.2101$ corresponding to the jump point G to evolve the probability over time. It can be clearly seen from figure 3(c) that the probabilities exhibit periodic oscillations, indicating that when the system parameters are taken from the continuous part of the type III boundary line, the system is stable. In figure 3(d), the probabilities increase exponentially with time, indicating that the system is unstable. This verifies our analysis. What needs to be particularly noted here is that in figure 3(a), the type III boundary line has gaps near $\gamma=0.5+n$ (see enlarged illustration) and $\gamma=n$ ($n=0,1,2,...$). When the parameters are taken within these gaps, Im($\rho$)$\neq0$ and the system is unstable, which is not shown here. At the same time, it can also be seen from figure 3(a) that for the system to undergo stable spin-flipping tunneling (corresponding to $\gamma=0.5+n$) or stable non-spin-flipping tunneling (corresponding to $\gamma=n$), the system parameters can only be taken from the type II boundary line.

\begin{figure}[htp]\flushleft
\includegraphics[height=1.7in,width=1.6in]{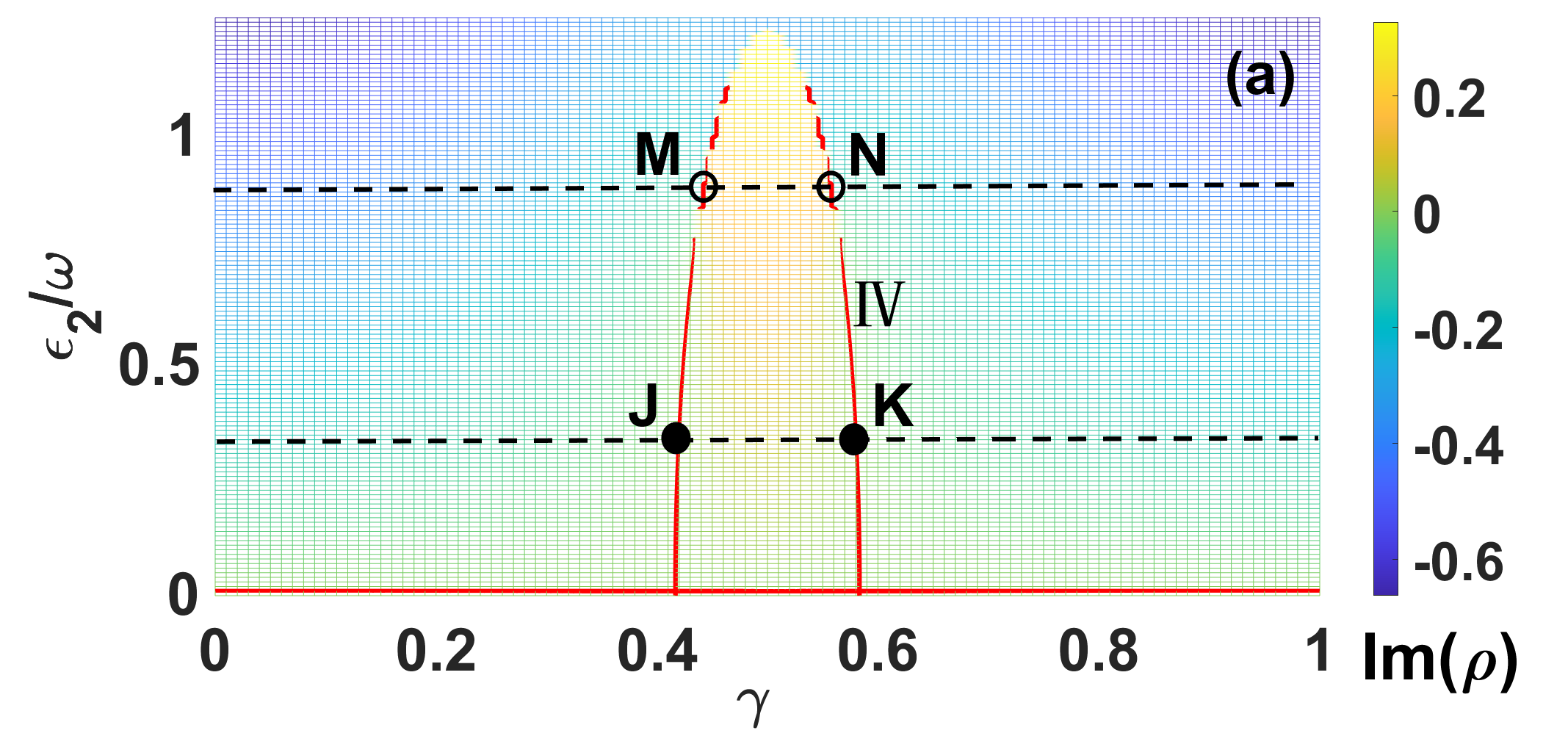}
\includegraphics[height=1.7in,width=1.6in]{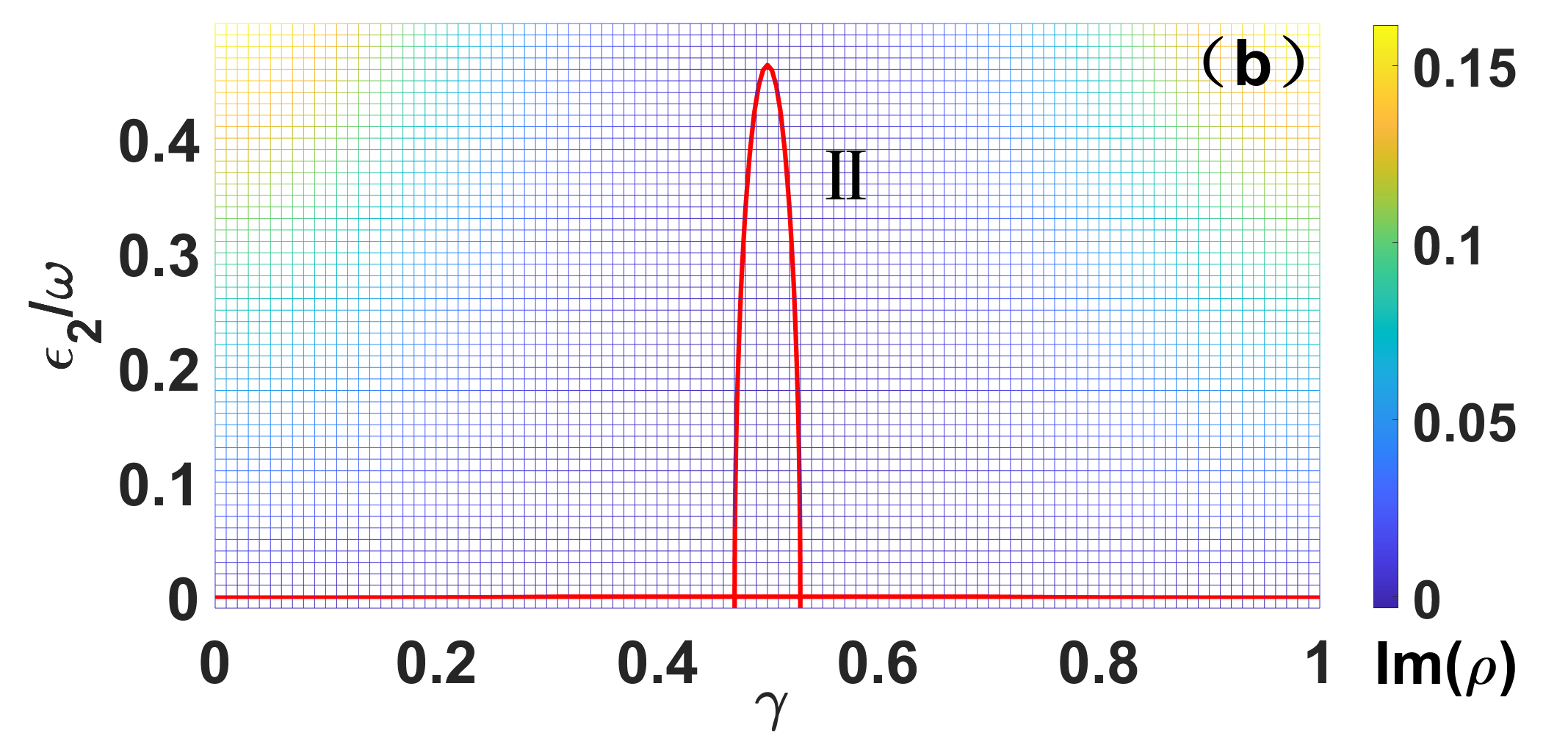}
\includegraphics[height=1.7in,width=1.5in]{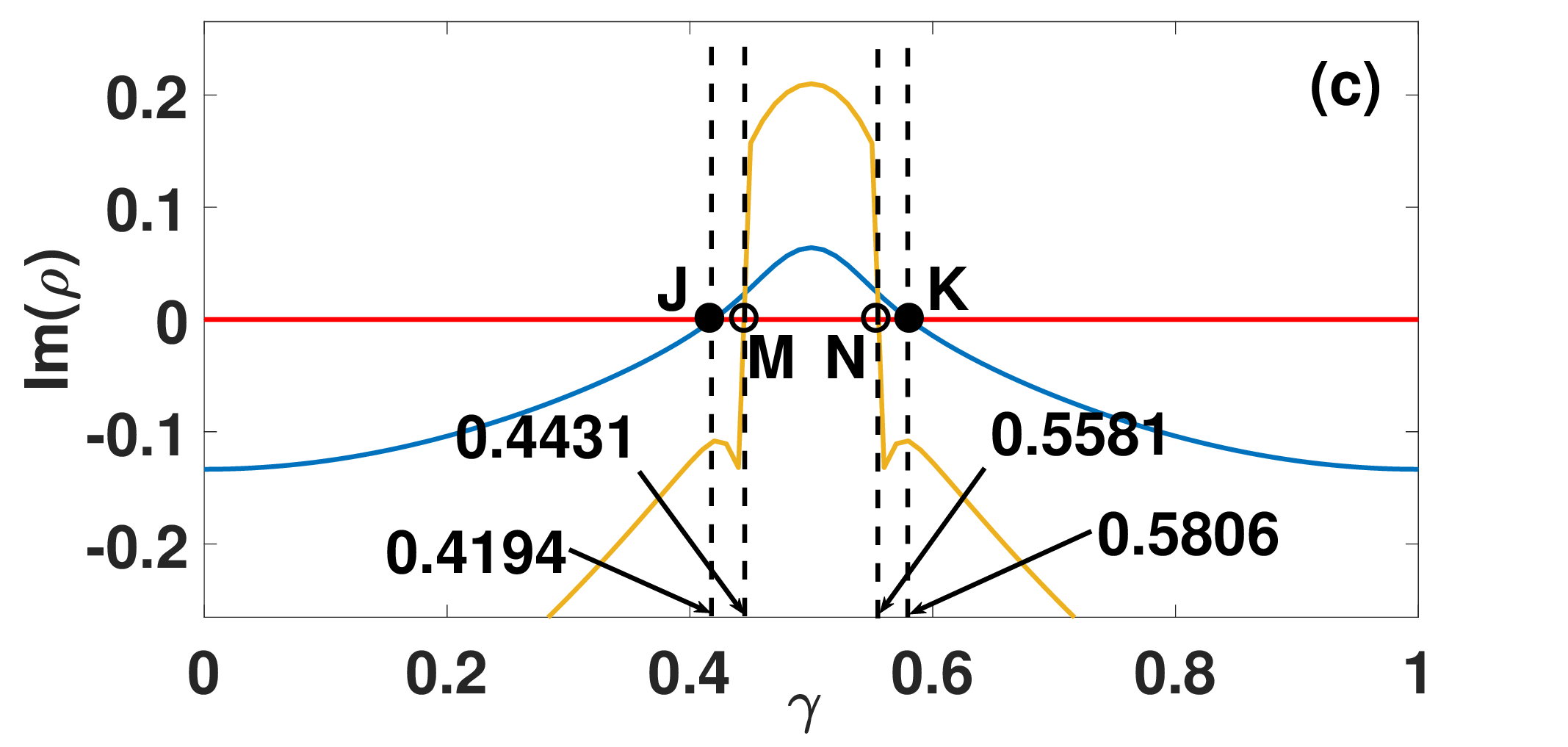}
\includegraphics[height=1.6in,width=1.6in]{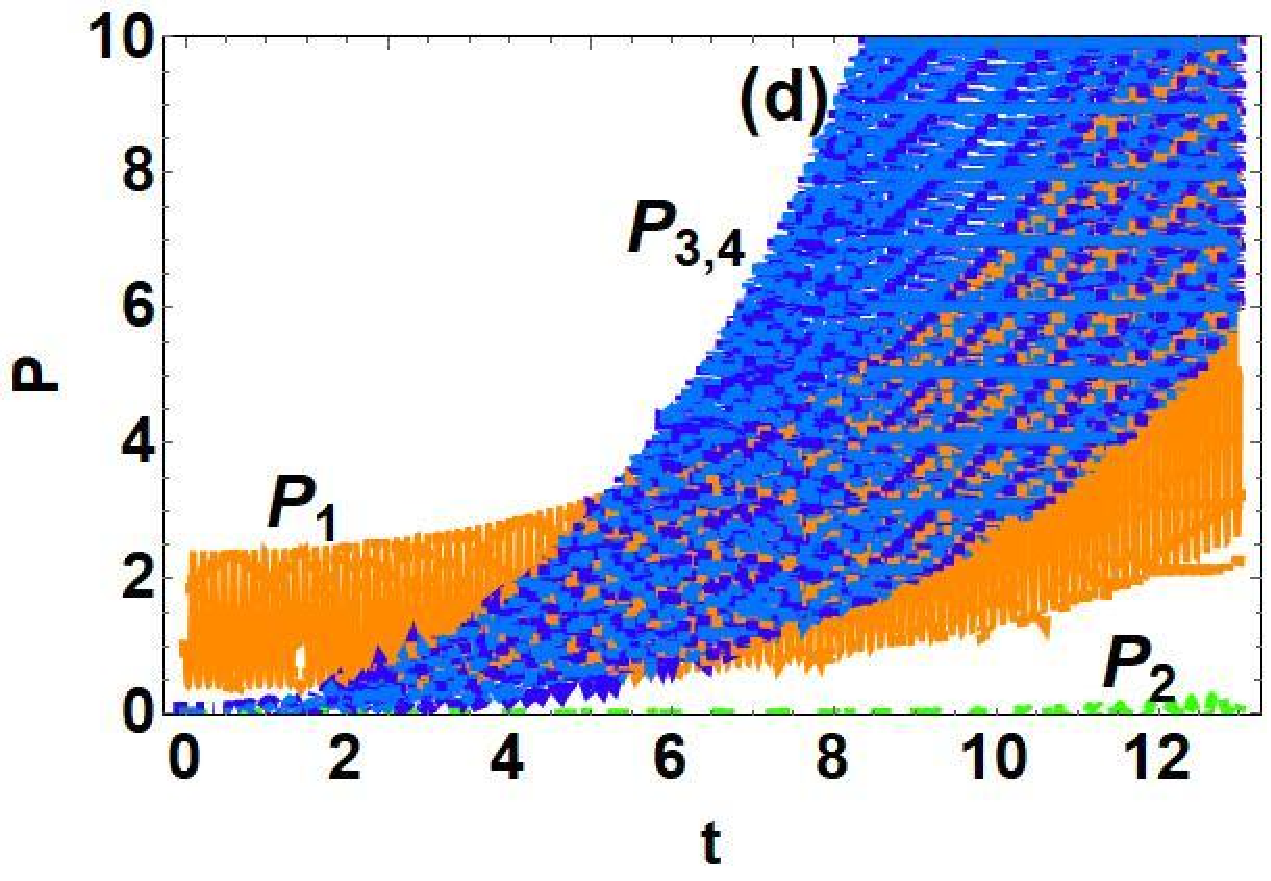}
\includegraphics[height=1.6in,width=1.4in]{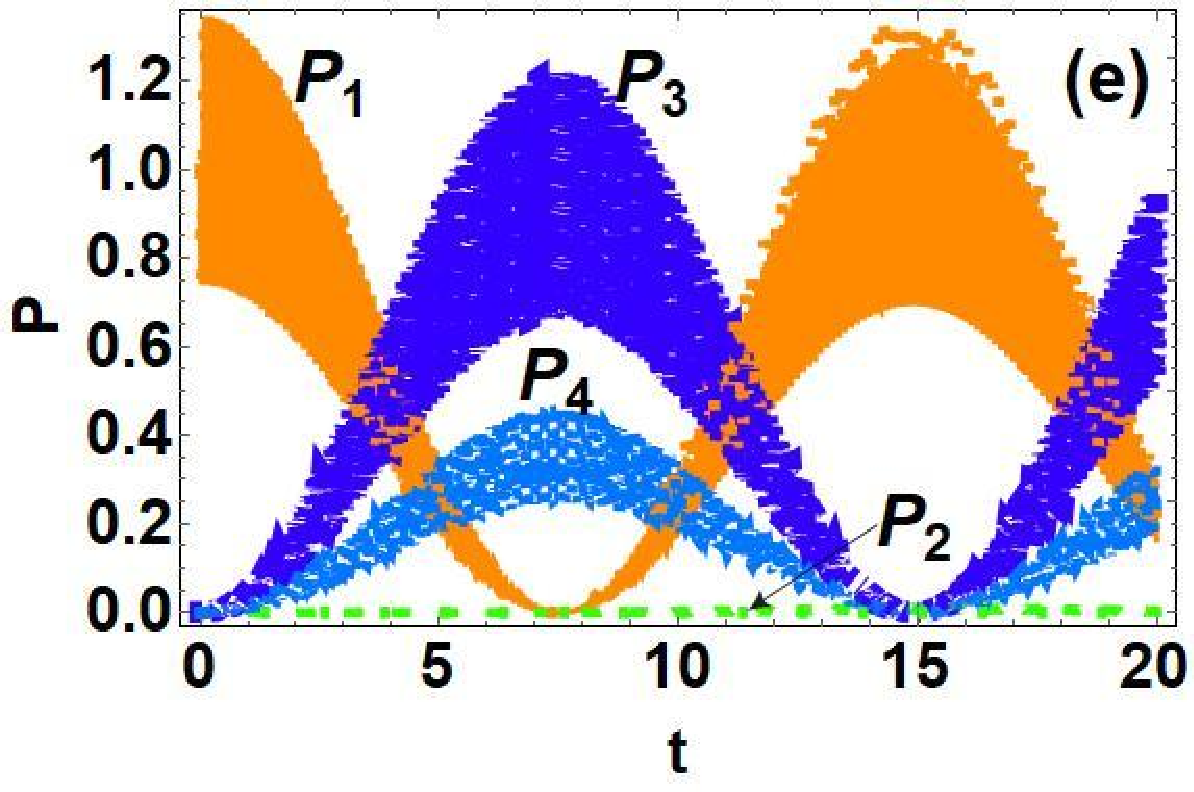}
\caption{\scriptsize{(a) and (b) Im($\rho$) as a function of $\gamma$ and $\frac{\epsilon_{2}}{\omega}$ for (a) $\frac{\epsilon_{1}}{\omega}=1$ and (b) $\frac{\epsilon_{1}}{\omega}=3.06$. (c) Im($\rho$) as a function of $\gamma$ for $\frac{\epsilon_{2}}{\omega}=0.88$(yellow solid line) and $\frac{\epsilon_{2}}{\omega}=0.3$(blue solid line). (d) and (e) show the time-evolution curves of probabilities $P_{k}=|a_{k}|^{2}$ for (d) $\gamma=0.4431$, $\frac{\epsilon_{2}}{\omega}=0.88$ and (e) $\gamma=0.4194$, $\frac{\epsilon_{2}}{\omega}=0.3$, starting the system with a spin-up particle in the right well. The other parameters are chosen as $\nu=1$, $\Omega=100$, and $\omega=50$.}}
\end{figure}

In figures 4(a) and 4(b), we respectively take the parameters $\frac{\epsilon_1}{\omega}=1$ and $\frac{\epsilon_1}{\omega}=3.06$ to plot Im($\rho$) as a function of $\gamma$ and $\frac{\epsilon_{2}}{\omega}$. In figure 4(b), the boundary line is the continuous boundary line of type II. Thus, when the parameters are taken on the continuous line, the system is stable, which has been verified in the previous part. Next, we will mainly discuss the boundary line of type IV in figure 4(a), which consists of both continuous and discontinuous boundary lines. The range of $\gamma$ corresponding to the discontinuous boundary line is approximately $\gamma\in(0.4315,0.5685)$. So, what are the differences in the stability of the system when the parameters are taken on the continuous and discontinuous parts of the type IV boundary line? Here, we take $\frac{\epsilon_2}{\omega}=0.88$ and $\frac{\epsilon_2}{\omega}=0.3$ as an example to draw horizontal dashed lines, which intersects the discontinuous and continuous parts of the type IV boundary line at four points M, N, J, and K, respectively, as shown in figure 4(a). Then, we set $\frac{\epsilon_2}{\omega}=0.88$ and $\frac{\epsilon_2}{\omega}=0.3$ to plot Im($\rho$) as a function of $\gamma$ in figure 4(c). It can be clearly seen from figure 4(c) that the points M and N on the discontinuous part of the type IV boundary line are jump points, with Im($\rho$) $\neq0$, and the system is unstable. However, the points J and K on the continuous part of the type IV boundary line are continuous points, with Im($\rho$) $=0$, and the system is stable. To verify our analysis, we separately take the parameters $\gamma=0.4431$ and $\frac{\epsilon_2}{\omega}=0.88$ corresponding to the jump point M, and take the parameters $\gamma=0.4194$ and $\frac{\epsilon_2}{\omega}=0.3$ corresponding to the continuous point J to evolve the probability over time, as shown in figures 4(d) and 4(e). Obviously, the probabilities increase exponentially in figure 4(d), indicating that the system is unstable. In contrast, figure 4(e) shows periodic oscillations of the probabilities, indicating that the system is stable. This is consistent with our previous analysis.

\subsection{Odd $\Omega$/$\omega$}

When $\Omega$/$\omega$ is odd, the Floquet quasienergies in equation (7) reduce to $E_{1,4}=\pm\rho_{-}$ and $E_{2,3}=\mp\rho_{+}$ with $\rho_{\pm}=J_{\frac{\Omega}{\omega}}\pm \sqrt{J_{0}^{2}}$. The stability of the system depends on Im($\rho_{-}$) and Im($\rho_{+}$). Below, we will also discuss the impact of parameters on system stability from two scenarios: scenario (i) and scenario (ii).

1. Scenario (i)

For odd $\Omega$/$\omega$ and $\epsilon_{1}= 0$ and $\epsilon_{2}> 0$, $\mathcal{J}_{0}(\frac{i\epsilon_{2}}{\omega})>1$ and $\mathcal{J}_{\frac{\Omega}{\omega}}(\frac{i\epsilon_{2}}{\omega})$ is a pure imaginary number. In this case, we are surprised to find that in order to make all the quasienergies real, $\gamma$ must be an integer, namely, $\gamma=0,1,2,...$.
At this time, $J_{\frac{\Omega}{\omega}}=\nu\sin(\pi\gamma) \mathcal{J}_{\frac{\Omega}{\omega}}(\frac{i\epsilon_{2}}{\omega})=0$ and $\rho_{\pm}=\pm \sqrt{J_{0}^{2}}$. The quasienergies are correspondingly reduced to $E_{1}=E_{2}=-E_{3}=-E_{4}=-\sqrt{J_{0}^{2}}$, which are all real. Because of $\gamma=0,1,2,...$, only stable quantum tunneling without spin-flipping can occur, while stable quantum tunneling with spin-flipping cannot happen.

2. Scenario (ii)

For odd $\Omega$/$\omega$ and $\epsilon_{1}> 0$ and $\epsilon_{2}> 0$, $\mathcal{J}_{0}(\frac{\epsilon_{1}+i\epsilon_{2}}{\omega})$ and $\mathcal{J}_{\frac{\Omega}{\omega}}(\frac{\epsilon_{1}+i\epsilon_{2}}{\omega})$ are  generally complex numbers. Therefore, it is very difficult to find appropriate parameter values that make both $\rho_{-}$ and $\rho_{+}$ real numbers. However, when $\gamma =n$ or $\gamma =0.5+n$ ($n=0,1,2,\ldots$), $\rho_{\pm}=\pm \sqrt{J_{0}^{2}}$ or $\rho_{\pm}=J_{\frac{\Omega}{\omega}}$, respectively. Under these conditions, it may be possible to find suitable parameters that simultaneously satisfy Im($\rho_{-}$)$=0$ and Im($\rho_{+}$)$=0$, in which case the quasienergies of the system are all real and the system is stable. In figures 5(a) and 5(b), we set $\gamma=0.5$ and $\gamma=1$ to plot Im($\rho_{+}$) as a function of $\frac{\epsilon_{1}}{\omega}$ and $\frac{\epsilon_{2}}{\omega}$, respectively. From figure 5(a), it can be seen that only the boundary line of type II exists. According to the previous analysis, when the parameters are taken on this type of continuous boundary line, the stable spin-flipping tunneling can happen. From figure 5(b), we can see that there are two types of boundary lines, namely, the discontinuous boundary line of type I and the continuous boundary line of type II. The stable tunneling without spin-flipping can only occur when the parameters are taken on the continuous boundary line of type II. They are not shown here.

\begin{figure}[htp]
\includegraphics[height=1.6in,width=1.6in]{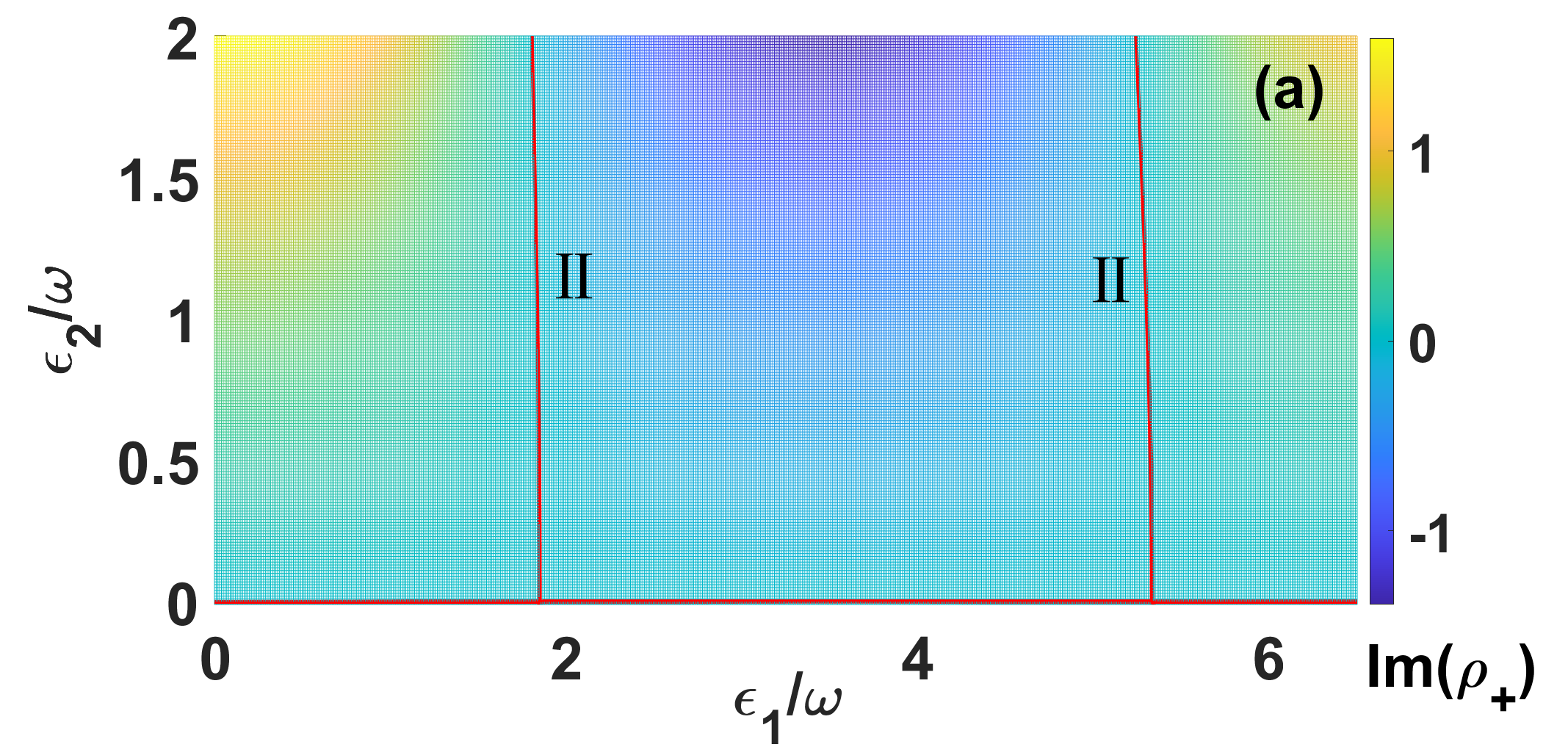}
\includegraphics[height=1.6in,width=1.6in]{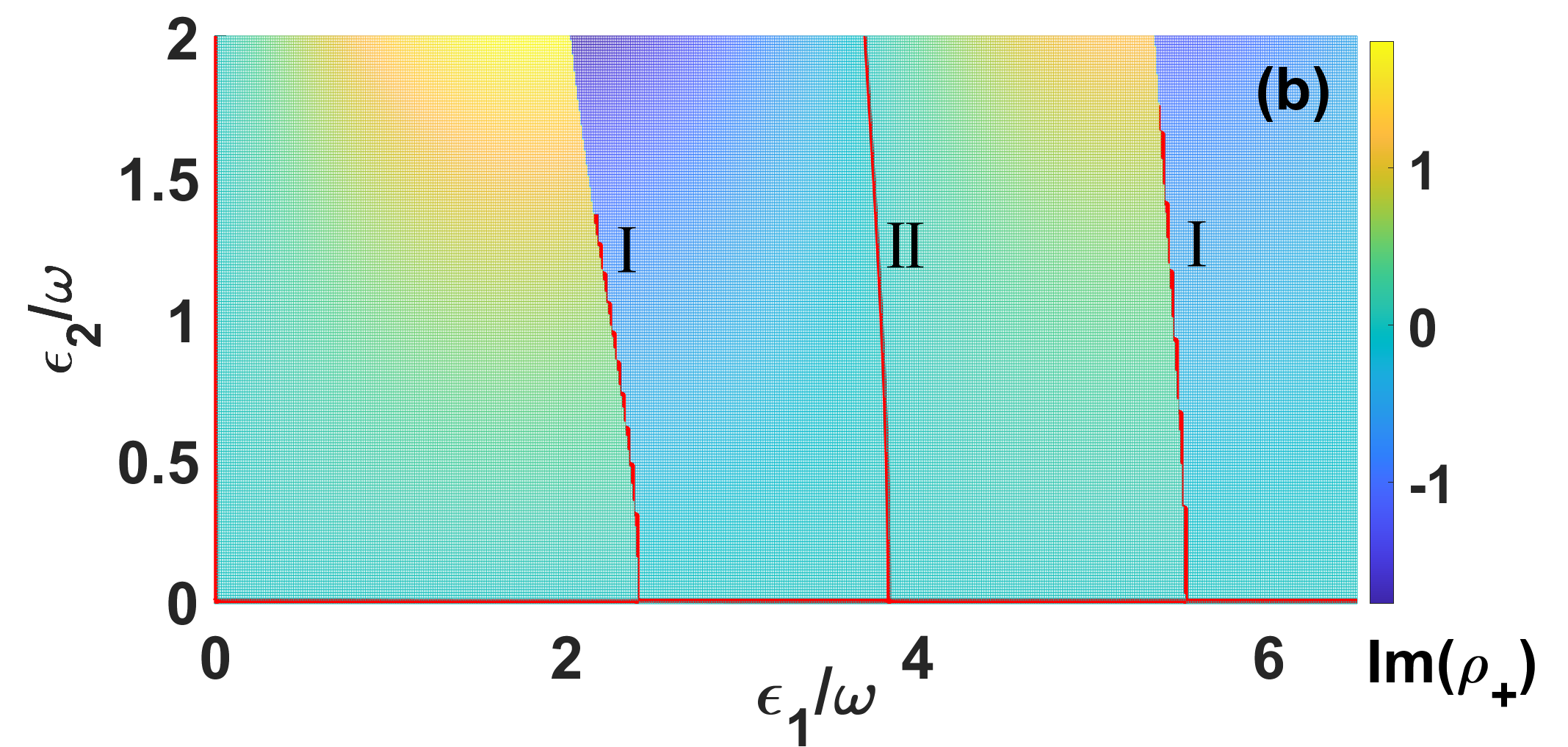}
\caption{\scriptsize{Im($\rho_{+}$) as a function of $\frac{\epsilon_{1}}{\omega}$ and $\frac{\epsilon_{2}}{\omega}$ for different SO coupling strengths:
(a) $\gamma=0.5$ and (b) $\gamma=1$. The other parameters are chosen as $\nu=1$, $\Omega=50$, and $\omega=50$.}}
\end{figure}

\section{CONCLUSION}

In summary, we have studied the stability of spin quantum dynamics for a SO coupled boson trapped in a driven non-Hermitian double well. By use of high-frequency approximation, the Floquet states and complex quasienergies of the system, as well as the non-Floquet states, have been obtained. We unexpectedly find that when $\Omega$/$\omega$ is even, for the case where the bias field strength is zero, the spin dynamics of the system are stable across the entire parameter region, and the quantum tunneling without spin-flipping is faster than that with spin-flipping. For the case where the bias field strength is not zero, the system's spin dynamics are unstable when the parameters lie on the discontinuous boundary line of type I and the discontinuous parts of the boundary lines of type III and type IV, but are stable when the parameters lie on the continuous boundary line of type II and the continuous parts of the boundary lines of type III and type IV. It is worth noting that when the parameters are chosen appropriately, the system can remain stable for arbitrary SO coupling strength, see figure 3(a). When $\Omega$/$\omega$ is odd, the SO coupling strength can only take integer values for the system's quantum dynamics to be stable if the bias field strength is zero. In this case, only non-spin-flipping stable tunneling can occur. Conversely, if the bias field strength is not zero, the stability is achieved only when the SO coupling strength takes integer or half-integer values and all other parameters reside on the continuous boundary line of type II, see figure 5. These findings establish an adjustable parity-governed mechanism for stabilizing spin dynamics and expand the possibilities for controlling stable spin dynamics in a non-Hermitian SO coupled system.

\section*{ACKNOWLEDGMENTS}

This work was supported by the National Natural Science Foundation of China under Grant No. 11747034, the Hunan Provincial Natural Science Foundation of China under Grant No. 2021JJ30435,
the Youth Student Basic Research Project of Natural Science Foundation of Hunan Province under Grant No. 2024JJ10039, the Scientific Research Foundation of Ningxia Education Department under Grant No. NYG2024202, National Students' Platform for Innovation and Entrepreneurship Training Program under Grants No. S202410542012 and No. 202510542020, and "Tenglong" Innovative Talent Fund of Hunan Normal University under Grant No. 2025TL104.

Z. D. Luo and Y. R. Yang contributed equally to this work.


\begin{thebibliography}{999}

\bibitem{moise} Moiseyev N 2011 Non-Hermitian quantum mechanics \emph{Cambridge: Cambridge University Press}
\bibitem{gana14} EI-Ganainy R, Makris K, Khajavikhan M, Musslimani Z, Rotter S and Christodoulides D 2018 Non-Hermitian physics and PT symmetry \emph{Nat. Phys.} 14 11
\bibitem{ashida} Ashida Y, Gong Z and Ueda M 2020 Non-Hermitian physics \emph{Adv. Phys.} 69 249
\bibitem{okuma} Okuma N and Sato M 2022 Non-Hermitian topological phenomena: a review \emph{arXiv:} 2205.10379
\bibitem{zhang} Zhang Y and Wei Z 2025 Non-Hermitian skin effect in non-Hermitian optical systems \emph{Laser Photonics Rev.} 19 2400099
\bibitem{du134} Wu Y, Zhu D, Wang Y, Rong X and Du J 2025 Experimental observation of Dirac exceptional points \emph{Phys. Rev. Lett.} 134 153601
\bibitem{miri} Miri M and Al\`{u} A 2019 Exceptional points in optics and photonics \emph{Science} 363 eaar7709
\bibitem{chen134} Cai Z, Li H and Chen W 2025 Quantum-classical correspondence of non-Hermitian symmetry breaking \emph{Phys. Rev. Lett.} 134 240201
\bibitem{gong91} Gong J and Wang Q 2015 Stabilizing non-Hermitian systems by periodic driving \emph{Phys. Rev. A} 91 042135
\bibitem{luoxb94} Yang B, Luo X, Hu Q and Yu X 2016 Exact control of parity-time symmetry in periodically modulated nonlinear optical couplers \emph{Phys. Rev. A} 94 043828
\bibitem{luoxb95} Luo X, Yang B, Zhang X, Li L and Yu X 2017 Analytical results for a parity-time symmetric two-level system under synchronous combined modulations \emph{Phys. Rev. A} 95 052128
\bibitem{zhou91} Zhou Z, Zhu B, Wang H and Zhong H 2020 Stability and collisions of quantum droplets in $\mathcal{PT}$-symmetric dual-core couplers \emph{Commun. Nonlinear Sci. Numer. Simulat.} 91 105424
\bibitem{bender80} Bender C M and Boettcher S 1998 Real spectra in non-Hermitian Hamiltonians having $\mathcal{PT}$ symmetry \emph{Phys. Rev. Lett.} 80 5243
\bibitem{bender40} Bender C M, Boettcher S and Meisinger P N 1999 $\mathcal{PT}$-symmetric quantum mechanics \emph{J. Math. Phys.} 40 2201
\bibitem{xiao85} Xiao K, Hai W and Liu J 2012 Coherent control of quantum tunneling in an open double-well system \emph{Phys. Rev. A} 85 013410
\bibitem{lin471} Lin Y, Jim\'{e}nez-Garc\'{\i}a K and Spielman I B 2011 Spin-orbit-coupled Bose-Einstein condensates \emph{Nature} 471 83
\bibitem{kato306} Kato Y K, Myers R C, Gossard A C and Awschalom D D 2004 Coherent spin manipulation without magnetic fields in strained semiconductors \emph{Science} 306 1910
\bibitem{bern314} Bernevig B A, Hughes T L and Zhang S C 2006 Quantum Spin Hall Effect and Topological Phase Transition in HgTe Quantum wells \emph{Science} 314 1757
\bibitem{saka18} Sakaguchi H and Malomed B 2016 One- and two-dimensional solitons in PT-symmetric systems emulating spin-orbit coupling \emph{New J. Phys.} 18 105005
\bibitem{qin24} Qin J, Zhou L and Dong G 2022 Imaginary spin-orbit coupling in parity-time symmetric systems with momentum-dependent gain and loss \emph{New J. Phys.} 24 063025
\bibitem{zhao108} Zhao X and Zhou L 2023 Stable molecular state under dissipative spin-orbit coupling \emph{Phys. Rev. A} 108 013311
\bibitem{liu109} Liu D, Ren Z, Wong W, Zhao E, He C, Pak K, Jo G and Li J 2024 Complete interband transitions for non-Hermitian spin-orbit-coupled cold-atom systems \emph{Phys. Rev. A} 109 053305
\bibitem{xu65} Xu Z, Zhou Z, Cheng E, Lang L and Zhu S 2022 Gain/loss effects on spin-orbit coupled ultracold atoms in two-dimensional optical lattices \emph{Sci. China-Phys. Mech. Astron.} 65 283011
\bibitem{luo22} Luo Y, Wang X, Luo Y, Zhou Z, Zeng Z and Luo X 2020 Controlling stable tunneling in a non-Hermitian spin-orbit coupled bosonic junction \emph{New J. Phys.} 22 093041
\bibitem{tang55} Tang J, Hu Z, Zeng Z, Xiao J, Li L, Chen Y, Chen A and Luo X 2022 Spin Josephson effects of spin-orbit-coupled Bose-Einstein condensates in a non-Hermitian double well \emph{J. Phys. B} 55 245301
\bibitem{xie56} Xie X, Cui J, Luo Z, Xie Y, Li W, Hai W and Luo Y 2023 Analytical results for a spin-orbit coupled atom held in a non-Hermitian double well under synchronous combined modulations \emph{J. Phys. A} 56 505302
\bibitem{luoxb103} Luo X, Zeng Z, Guo Y, Yang B, Xiao J, Li L, Kong C and Chen A 2021 Controlling directed atomic motion and second-order tunneling of a spin-orbit-coupled atom in optical lattices \emph{Phys. Rev. A} 103 043315
\bibitem{zou17} Zou M, Lu G, Luo Y and Hai W 2020 Quantum transport and control of a classically chaotic open system \emph{Results Phys.} 17 103157
\bibitem{li10} Li J, Harter A K, Liu J, Melo L, Joglekar Y N, and Luo L 2019 Observation of parity-time symmetry breaking transitions in a dissipative Floquet system of ultracold atoms \emph{Nat. commun.} 10 855
\bibitem{xue90} Yu Z and Xue J 2014 Selective coherent spin transportation in a spin-orbit-coupled bosonic junction \emph{Phys. Rev. A} 90 033618
\bibitem{luo93} Luo Y, Lu G, Kong C and Hai W 2016 Controlling spin-dependent localization and directed transport in a bipartite lattice \emph{Phys. Rev. A} 93 043409
\bibitem{luo110} Luo X, Huang J, Zhong H, Qin X, Xie Q, Kivshar Y, and Lee C 2013 Pseudo-Parity-Time Symmetry in Optical Systems \emph{Phys. Rev. Lett.} 110 243902
\bibitem{longhi80} Longhi S 2009 Bloch oscillations and Wannier-Stark localization in a tight-binding lattice with increasing intersite coupling \emph{Phys. Rev. B} 80 033106
\bibitem{kfir123} Kfir O 2019 Entanglements of Electrons and Cavity Photons in the Strong-Coupling Regime \emph{Phys. Rev. Lett.} 123 103602
\bibitem{fel124} Felicetti S and Boit\'{e} A L 2020 Universal Spectral Features of Ultrastrongly Coupled Systems \emph{Phys. Rev. Lett.} 124 040404
\bibitem{shirley138} Shirley J H 1965 Solution of the Schr\"{o}dinger equation with a Hamiltonian periodic in time \emph{Phys. Rev.} 138 B979
\bibitem{sambe7} Sambe H 1973 Steady states and quasienergies of a quantum-mechanical system in an oscillating field \emph{Phys. Rev. A} 7 2203
\bibitem{jinas322} Jinasundera T, Weiss C and Holthaus M 2006 Many-particle tunnelling in a driven Bosonic Josephson junction \emph{Chem. Phys.} 322 118
\bibitem{lu83} Lu G, Hai W and Xie Q 2011 Coherent control of atomic tunneling in a driven triple well \emph{Phys. Rev. A} 83 013407
\bibitem{guo103} Guo A, Salamo G, Duchesne D, Morandotti R and Christodoulides D 2009 Observation of PT-symmetry breaking in complex optical potentials \emph{Phys. Rev. Lett.} 103 093902
\bibitem{uch68} Uchiyama C and Aihara M 2003 Synchronized pulse control of decoherence \emph{Phys. Rev. A} 68 052302




















\end{thebibliography}
\end{document}